\documentclass{aa}
\usepackage{graphicx}
\usepackage{rotating}
\usepackage{txfonts}

\def\dsct{$\delta$~Scuti }
\def\gdor{$\gamma$~Doradus }
\def\bv0{$(B-V)_{0}$}
\def\pms{pre-main sequence }

\def\Msun{$M_{\odot}$ }
\def\Lsun{$L_{\odot}$}

\def\Teff{$T_{\rm eff}$ }

\def\vsini{$v\cdot\sin{i}$ }

\begin{document}

\title{Pulsating pre-main sequence stars in IC 4996 and NGC 6530}

\author{K. Zwintz\inst{1}
\and
W. W. Weiss\inst{1}
}

\titlerunning{Pulsating PMS stars in IC 4996 and NGC 6530}

\offprints{K. Zwintz, \\ \email{zwintz@astro.univie.ac.at}}

\institute{Institute of Astronomy, University of Vienna,
T\"urkenschanzstrasse 17, A-1180 Vienna, Austria }

\date{Received / Accepted }

\abstract
{Asteroseismology of pulsating \pms (PMS) stars has the potential of testing the validity of current models of PMS structure and evolution. As a first step, a sufficiently large sample of pulsating PMS stars has to be established, which allows to select candidates optimally suited for a detailed asteroseismological analysis based on photometry from space or ground based network data.}
{A search for pulsating PMS members in the young open clusters IC 4996 and NGC 6530 has been performed to improve the sample of known PMS pulsators. As both clusters are younger than 10 million years, all members with spectral types later than A0 have not reached the zero-age main sequence yet. Hence, IC 4996 and NGC 6530 are most suitable to search for PMS pulsation among their A- and F-type cluster stars.}
{CCD time series photometry in Johnson $B$ and $V$ filters has been obtained for IC 4996 and NGC 6530. The resulting light curves for 113 stars in IC 4996 and 194 stars in NGC 6530 have been subject to detailed frequency analyses.}
{2 $\delta$ Scuti-like PMS stars have been discovered in IC 4996 and 6 in NGC 6530. For another PMS star in each cluster, pulsation can only be suspected. According to the computed pulsation constants, the newly detected PMS stars seem to prefer to pulsate in a similar fashion to the classical $\delta$ Scuti stars, and with higher overtone modes.}
{}


\maketitle

\section{Introduction}

The study of the first stages in the formation of stars is currently one of the most active research fields in stellar astronomy.
Pre-main sequence (PMS) stars are contracting from the birthline to the zero-age main sequence (ZAMS) in the Hertzsprung-Russell (HR) diagram.
They interact with the circumstellar environment in which they are still embedded; hence they are characterized by a large degree of activity, strong near- or far-IR excesses, and very often by emission lines. Two major groups can be distinguished: T Tauri and Herbig Ae/Be objects. Members of both groups show photometric and spectroscopic variabilities on timescales from minutes to years, indicating that stellar activity begins in the earliest phases of stellar evolution, prior to the arrival on the main sequence.

T Tauri stars are newly formed low-mass stars that have recently become visible in the optical range. As they have primarily spectral types ranging from late F to M, they are not the prime targets to search for PMS pulsation.
The Herbig Ae/Be (HAEBE) stars are intermediate-mass (1.5\,--\,10\,$M_{\odot}$) objects (e.g., Herbig  \cite{her60}; Finkenzeller \& Mundt \cite{fin84}), which are either in their PMS phase ($M < 6\,M_{\odot}$) or have already arrived on the ZAMS ($M \geq 6\,M_{\odot}$). They have spectral types B and A, are characterized by emission lines, and mostly show excess radiation in the infrared. PMS stars with masses between 1.5 and 4\,$M_{\odot}$ have the right combination of temperature, luminosity, and mass to become vibrationally unstable, and hence are the main targets in which to search for pulsation.

During their evolution to the main sequence, young stars move across the instability region in the HR diagram, which suggests that part of their activity is due to stellar pulsations. PMS stars differ from their counterparts with the same effective temperature, mass, and luminosity, but that have already evolved off the main sequence, mostly in the inner regions, although their atmospheres are quite similar (Marconi \& Palla \cite{mar98}). The discovery of pulsating PMS stars is important because it allows to constrain the internal structure of young stars and to test evolutionary models.

As the evolutionary tracks for pre- and post-main sequence stars intersect each other several times (e.g., Breger \& Pamyatnykh \cite{bre98}), the determination of the evolutionary stage of a field star may be ambiguous. Additional information, like the age or distance of the star, is needed to resolve this ambiguity.
Young clusters are most suitable to search for PMS pulsators because all members have the same age and distance, hence confusion with  more evolved objects can be reduced.

The first two \pms pulsators, V\,588\,Mon (HD\,261331, NGC\,2264\,2) and V\,589\,Mon (HD\,261446, NGC\,2264\,20), were discovered by Breger (\cite{bre72}) in the young open cluster NGC 2264. It took more than 20 years before the discovery of the next PMS pulsating star. Kurtz \& Marang (\cite{kur95}) observed pulsations in the well-studied Herbig Ae field star HR 5999 (HD\,144668, V856\,Sco). Since then, several detections of pulsations in Herbig Ae field stars (e.g., Donati et al. \cite{don97}; Marconi et al. \cite{mar00}; etc.), as well as in members of young clusters (e.g., Zwintz et al. \cite{zwi05}) have been made.

In Sect. 2, the general characteristics of the two open clusters IC 4996 and NGC 6530 are summarized; Sects. 3 and 4 describe the observations, data reduction, and frequency analysis. The pulsating PMS stars in IC 4996 and NGC 6530 are shown in Sects. 5 and 6, respectively, and the calculation of the according pulsation constants is given in Sect. 7. As all stars in the fields of the clusters were analyzed, the other detected variable stars are listed in Sect. 8. A summary of the cluster properties derived from our analysis can be found in Sect. 9, while the conclusions of this paper are specified in Sect. 10.

\section{The young open clusters IC 4996 and NGC 6530}
IC 4996 ($\alpha_{2000} = 20^{\rm h} 16\fm 5, \delta_{2000} = +37\degr 38\farcm 0$)
is located in the direction of Cygnus, 40 pc above the galactic plane, and is part of a large region with active star formation that contains other young open clusters and Wolf-Rayet stars. An IRAS map of the region (Lozinskaya \& Repin \cite{loz90}) shows the presence of a dusty shell that surrounds the cluster.

The age and distance estimates for IC 4996 from different authors (e.g., Alfaro et al. \cite{alf85}; Delgado et al. \cite{del98}) agree: the cluster is slightly younger than 10$^7$ years and is located $\sim$\,1.7 kpc from the Sun. In particular, the inferred age indicates the likely existence of \pms members in the range of spectral types A and F.
Delgado et al. (\cite{del99}) performed spectroscopic observations with the aim of estimating radial velocities and spectral types for 16 proposed PMS stars to confirm or reject their cluster membership. Additionally, the authors found spectral features indicative of the PMS nature of the stars with spectral types later than A0.

NGC 6530 ($\alpha_{2000} = 18^{\rm h} 04\fm 5, \delta_{2000} = -24\degr 22\farcm 0$) is located in the central part of the HII region M8, the Lagoon nebula. Since the first study performed by Trumpler in \cite{tru30}, several investigations have been devoted to the study of this cluster.
The authors agree to an age of 1.0 -- 3.0 Myr (Kilambi \cite{kil77}; Sung et al. \cite{sun00}), a distance of 1.86 $\pm$ 0.07 kpc (Mc Call \cite{mcc90}), and a color excess $E(B-V) = 0.39 \pm 0.09$ mag (Chini \& Neckel \cite{chi81}).

Kilambi (\cite{kil77}) obtained $UBV$ photographic photometry of NGC 6530 and found that all stars fainter than $V$\,=\,12.0\,mag have not reached the ZAMS yet.
Van den Ancker et al. (\cite{van97}) obtained Walraven {\it WULBV}, Johnson/Cousins {\it UBV(RI)}, and near-IR {\it JHK} photometric data, and performed spectroscopy of NGC 6530.

\section{Observations and data reduction}
\subsection{IC 4996}
IC 4996 was observed with the 1.5\,m telescope ($f/8$) at the Sierra Nevada Observatory, Spain, between Sept. 2 and Sept. 15, 2002, in Johnson $B$ \& $V$ filters using a 1k x 1k CCD chip with a scale of 0.33$\arcsec$/px providing a field of view of \mbox{6$\arcmin \times 6\arcmin$}. In total, 69.76 hours of time-series photometry have been obtained in 12 out of 14 granted nights. Only the first ten nights (corresponding to 62.38 hours of observations and 1990 science frames) were used for the analysis because during the last two nights the weather conditions were not good enough, resulting in an extremely high scatter in the light curves.

The images were already bias- and dark-corrected with the standard procedure used at the observatory.
The flat field correction was performed within the reduction software {\sc PODEX} written by Kallinger (\cite{kal05}). {\sc PODEX} allows to extract the photometric signal of CCD time series photometry, using a combination of aperture photometry and point-spread function fitting.
After the light curve is computed for each selected star, the mean value of the comparison light curve is subtracted, where the stars used for the comparison light curve can be selected arbitrarily. Optionally flat field and (color-dependent) extinction corrections can be applied. The output of the program not only contains the photometric signal for each integration, but also the values of the airmass and the comparison light curve. The advantage of {\sc PODEX} is that variable stars can be identified easily and immediately discarded for the comparison light curve.
This is specifically important if -- as in our case -- no a priori information about the variability of the analyzed stars is available.

\begin{figure}[htb]
\centering
\includegraphics[width=7.0cm]{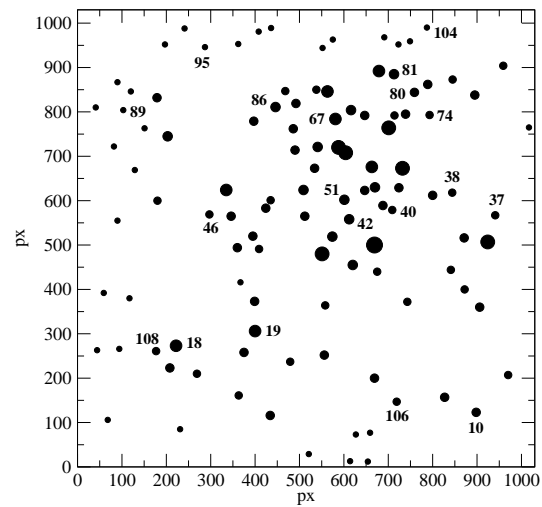}
\caption{Schematic map of the observed field of IC 4996 (fov 6$\arcmin \times 6\arcmin$) with all stars measured in Johnson {\it B} \& {\it V}, where 1 pixel (px) corresponds to 0.33$\arcsec$. Identifiers refer to the objects discussed in the text.}
\label{ic4996_chart}
\end{figure}

The 113 stars selected for the further analysis (see Fig. \ref{ic4996_chart}) lie in the range between $V\,=\,11\,-\,18$\,mag. Nightly means were subtracted to correct for zero-point changes and long-term irregular light variations, which most likely are due to variable extinction by circumstellar dust. Color-dependent extinction was taken into account, as described in Zwintz et al. (\cite{zwi05}).

Our own star numbers are used, but cross references with numbers used in the WEBDA database and given by Delgado et al. (\cite{del98} \& \cite{del99}) are listed, too.

\subsection{NGC 6530}
For NGC 6530, CCD photometric time series in Johnson {\it B} \& {\it V} filters were obtained with the 0.9\,m telescope ($f/13.5$) at the Cerro Tololo Interamerican Observatory (CTIO), Chile. Between Aug. 1--7 and Aug. 9--15, 2002, NGC 6530 was observed using the dedicated 2048 x 2046 SITe CCD chip, which provides a field of view of $\sim$\,13.5$\arcmin$\,$\times$\,13.5$\arcmin$ with a scale of 0.396$\arcsec$/px. In total, 80.16 hours of time-series photometry was acquired in 12 out of 14 nights. As the cluster was slightly too large to be observed on a single frame, two overlapping regions were chosen for which the observing time was shared. Out of 3437 scientific frames, 1601 were observed for field 1 and 1336 for field 2.

Bias subtraction and the creation of superflat images were performed using the {\tt IRAF ared.quad} package. Flat field correction was performed with the reduction software {\sc PODEX}, which was also used to extract the photometric signal, as it was for the data of IC 4996.

\begin{figure}[htb]
\centering
\includegraphics[width=6.5cm]{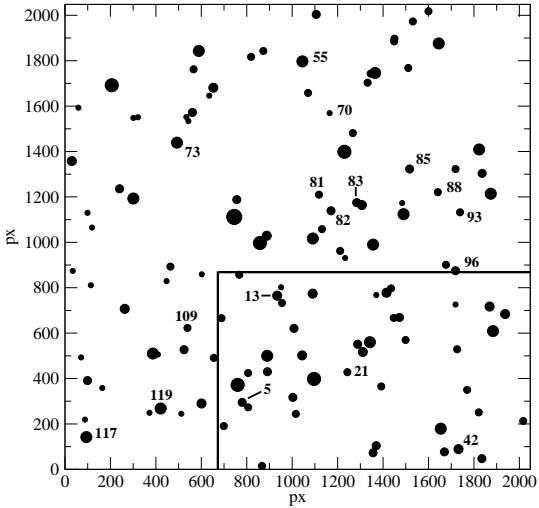}
\caption{Schematic map of the observed field 1 of NGC 6530 (fov\,$\sim$\,13.5\arcmin\,$\times$\,13.5\arcmin), South is at the top and East is to the left, with all stars measured in Johnson {\it B} \& {\it V}, where 1 pixel (px) corresponds to 0.369\,\arcsec. The area marked with the solid lines corresponds to the overlapping region. Identifiers refer to the objects discussed in the text.}
\label{ngc6530_1-chart}
\end{figure}

\begin{figure}[htb]
\centering
\includegraphics[width=6.5cm]{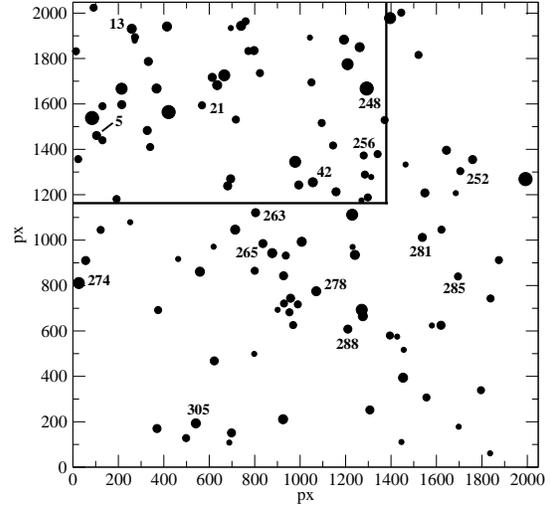}
\caption{Schematic map of the observed field 2 of NGC 6530, similar to Fig. \ref{ngc6530_1-chart}.}
\label{ngc6530_2-chart}
\end{figure}

In total, 194 stars have been identified in both fields where 43 stars lie in the overlapping region, 79 only in field 1 and 72 only in field 2 (see Figs. \ref{ngc6530_1-chart} and \ref{ngc6530_2-chart}). Nightly means were subtracted from the light curves to correct for zero-point changes and  long-term irregular light variations caused by variable extinction due to circumstellar material. Again, the effect of color-dependent extinction was encountered and corrected using the same method as for IC 4996 and NGC 6383 (Zwintz et al. \cite{zwi05}).

\section{Frequency analysis}
For all selected stars, detailed frequency analyses were performed using primarily {\sc Period98} (Sperl \cite{spe98}), which is based on the Discrete Fourier Transformation (DFT, Deeming \cite{dee75}) and provides a multi-sine fit option. Here, a signal was considered to be significant if it exceeds four times the noise level in the amplitude spectrum (Breger et al. \cite{bre93}; Kuschnig et al. \cite{kus97}).

Additionally, {\sc SigSpec} (Reegen \cite{ree06}), which computes significance levels for amplitude spectra of time-series with arbitrarily given sampling and white noise, has been used, providing a second independent method for detailed frequency analyses. {\sc SigSpec}'s advantage over classical techniques is the employment of analytically clean statistics instead of coarse signal-to-noise ratio estimations. A {\sc SigSpec}-significance of $\sim$5.46 corresponds to a signal-to-noise ratio of 4 (Breger et al. \cite{bre93}) mentioned before. Hence, a signal was considered to be significant within the {\sc SigSpec} analysis, if its significance was higher than 5.46.

Only if frequencies appeared significant in both filters and using both methods are they believed to be intrinsic and not artefacts of the reduction.

The errors of frequencies and amplitudes (Table \ref{ic4996-fa}) were calculated using the relations given by Montgomery (\cite{mon99}):

\begin{equation}
\sigma(a) = \sqrt{\frac{2}{N}}\,\cdot\,\sigma(m)
\label{a-error}
\end{equation}

\begin{equation}
\sigma(f) = \sqrt{\frac{6}{N}}\,\cdot\,\frac{1}{\pi\,T}\,\cdot\,\frac{\sigma(m)}{a},
\label{f-error}
\end{equation}

where $\sigma(m)$ is the rms magnitude of the data set, $a$ the corresponding amplitude, $N$ is the number of data points and $T$ is the time base of the observations.
It has to be noted that the errors calculated with the above formulae can only provide a lower limit of the true errors. If the errors in magnitude are correlated in time and the correlation time is of the order of the period of the signal, the error estimates are too low by a factor $\sqrt{A}$, where the function $A \approx 0.24 P / \Delta t$ with the separation of data points of $\Delta t$ (Montgomery \cite{mon99}).

\section{Pulsating PMS stars in IC 4996}
40 PMS cluster members have been selected as prime candidates for pulsation in IC 4996 due to their position in the HR diagram, where the major criterion has been the location of the stars in the region of the classical instability strip (e.g., Pamyatnykh \cite{pam00}).
Two bona fide, IC 4996 37 and 40, and one suspected, IC 4996 46, pulsating \pms stars have been discovered in this cluster. The previously known parameters of these stars are given in Table \ref{ic4996-pars} and their corresponding frequencies and amplitudes are listed in Table \ref{ic4996-fa}.

\begin{table}[htb]
\caption{Parameters of the bona fide and suspected PMS pulsators in IC 4996.}
\label{ic4996-pars}
\begin{center}
\begin{scriptsize}
\begin{tabular}{ccccccc}
\hline
Star & WEBDA & Delgado no. & $V$ & $(B-V)$ & $(U-B)$ & Sp\\
& & \# & [mag] & [mag] & [mag] & \\
\hline
37 & 201 & 32 & 15.30 & 0.80 & 0.44 & A5 \\
40 & 171 & 30 & 15.03 & 0.75 & 0.43 & A4 \\
46 & 1085 & 85 & 15.30 & 0.72 & 0.36 & - \\
\hline
\end{tabular}
\end{scriptsize}
\end{center}
\end{table}

\subsection{IC 4996 37}

For IC 4996 37, Delgado et al. (\cite{del98}) measured $V$\,=\,15.30\,mag, $(B-V)$\,=\,0.8\,mag, and $(U-B)$\,=\,0.44\,mag using CCD photometry. This coincides well with the observations performed by Vansevicius et al. (\cite{van96}), who give $V$\,=\,15.207\,mag and $(B-V)$\,=\,0.784\,mag, but also $(V-R)$\,=\,0.428\,mag and $(V-I)$\,=\,0.921\,mag. Delgado et al. (\cite{del99}) obtained long-slit spectra for 16 stars of IC 4996, among those also star 37, which was found to have a spectral type of A5. Hence, it is located in the region of the HR diagram, where pulsation can be expected, and this made it an ideal target to search for pulsation.

\begin{figure}[htb]
\centering
\includegraphics[width=8.5cm]{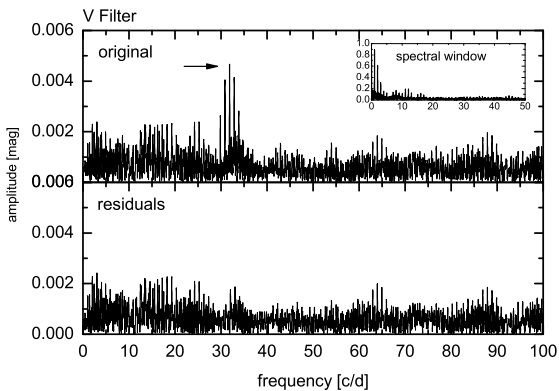}
\caption{Amplitude spectrum of IC 4996 37 in the $V$ filter; the identified frequency is marked with an arrow.}
\label{ic4996-37amps}
\end{figure}

The frequency analysis with {\sc Period98} yielded a single intrinsic significant frequency in both filters at 31.875\,c/d corresponding to a period of 45 minutes. This frequency was also detected using {\sc SigSpec} with significances of 14.7 in $V$ and 5.9 in $B$.
In the amplitude spectra of both filters, the peak at 31.875\,c/d can clearly be noticed; as an example, Fig. \ref{ic4996-37amps} shows the amplitude spectrum of the star in the $V$ filter. Several frequencies related to the one-day alias appear to be significant in the amplitude spectra of both filters. However they were omitted in the analysis, as they result from the daily gaps in the observations due to the day-night-cycle (see the appropriate spectral window).

\subsection{IC 4996 40}
For IC 4996 40, Delgado et al. (\cite{del98}) give $V$\,=\,15.03\,mag, $(B-V)$\,=\,0.75\,mag, and $(U-B)$\,=\,0.43\,mag. A spectral type of A4 was determined by Delgado et al. (\cite{del99}), which places the star inside the instability region in the HR diagram.

\begin{figure}[htb]
\centering
\includegraphics[width=8.5cm]{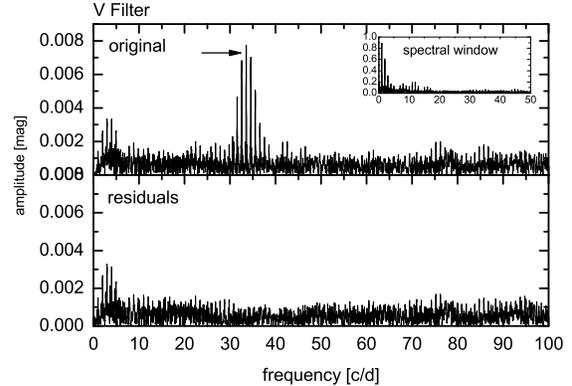}
\caption{Amplitude spectrum of IC 4996 40 in the $V$ filter; the identified frequency is marked with an arrow.}
\label{ic4996-40amps}
\end{figure}

The frequency analysis with {\sc Period98} resulted in the detection of a single significant frequency in both filters at 33.569\,c/d corresponding to a period of 43 minutes (Fig. \ref{ic4996-40amps}). The same frequency was also detected using {\sc SigSpec} with significances of 33.1 in $V$ and 22.2 in $B$. Again, a number of frequencies related to the one-day alias appear in the amplitude spectra and have been omitted in the analysis.

\subsection{IC 4996 46}
With $(B-V)$\,=\,0.72\,mag and $V$\,=\,15.3\,mag (Delgado et al. \cite{del98}), IC 4996 46 is also located in the region of the classical instability strip in the HR diagram. The frequency analysis resulted in the detection of different periods in $V$ and $B$ filters: two periods of approximately 6.0 and 6.5 hours with amplitudes of 5.4 and 3.8\,mmag in $V$ and two periods of 4.8 and 3.7 hours with amplitudes of 4.2\,mmag each in $B$. Hence, the star is considered as a suspected PMS pulsator.

\begin{table}[htb]
\caption{Frequencies and amplitudes determined for the two bona fide PMS pulsators in IC 4996, where the errors in the last digits of the corresponding quantities are given in parentheses.}
\label{ic4996-fa}
\begin{center}
\begin{tabular}{ccrrrrr}
\hline
Star & No. & \multicolumn{1}{c}{Frequency} & \multicolumn{1}{c}{{\it V} amp.} & \multicolumn{1}{c}{{\it B} amp.} \\
& & \multicolumn{1}{c}{[d$^{-1}$]} & \multicolumn{1}{c}{[mmag]} & \multicolumn{1}{c}{[mmag]} \\
 \hline
37 & f1 & 31.875(9) & 4.6(5) & 5.1(9) \\
40 & f1 & 33.569(7) & 7.6(5) & 8.5(7)  \\
\hline
\end{tabular}
\end{center}
\end{table}

\section{Pulsating PMS stars in NGC 6530}
30 PMS cluster members have been selected as prime candidates for pulsation in NGC 6530 according to their position in the HR diagram. Again, the major criterion has been the location of the stars in the region of the classical instability strip.
Six PMS pulsators were discovered with classical \dsct type frequencies (NGC 6530 5, 82, 85, 263, 278, and 281), and for a seventh star, pulsation can only be suspected (NGC 6383 288).

Table \ref{ngc6530_pulslist} shows an overview of all new pulsating PMS stars in NGC 6530, where $V$, $(B-V)$, and $(U-B)$ are taken from the WEBDA database, and the spectral types were derived by van den Ancker et al. (\cite{van97}).

\begin{table}[htb]
\caption{Parameters of the bona fide and suspected PMS pulsators in NGC 6530.}
\label{ngc6530_pulslist}
\begin{center}
\begin{footnotesize}
\begin{tabular}{ccccccc}
\hline
Star & WEBDA  & $V$ & $(B-V)$ & $(U-B)$ & Sp\\
&  & [mag] & [mag] & [mag] & \\
\hline
5 & 159  & 13.59 & 0.43 & 0.26 & -  \\
82 & 78  & 13.97 & 0.61 & 0.30 & - \\
85 & 53 & 13.07 & 0.65 & 0.35 & A1 III \\
263 & 57  & 13.67 & 0.63 & 0.35 & - \\
278 & 38  & 12.17 & 0.53 & 0.36 & A0/A5 \\
281 & 13  & 13.35 & 0.45 & 0.27 & - \\
288 & 28  & 13.23 & 0.41 & 0.35 & - \\
\hline
\end{tabular}
\end{footnotesize}
\end{center}
\end{table}

\subsection{NGC 6530 5}
NGC 6530 5 is situated in the region where the two observed fields overlapped.
For star 5 (WEBDA \#159), no information about its spectral type or cluster membership is available, but $UBV$ CCD photometry gives $V$\,=\,13.59\,mag, $(B-V)$\,=\,0.43\,mag, and $(U-B)$\,=\,0.26\,mag (Sung et al. \cite{sun00}). Hence, from the location in the cluster HR diagram, it is most likely a \pms member lying in the region of the classical instability strip.

\begin{figure}[htb]
\centering{\includegraphics[width=8.5cm]{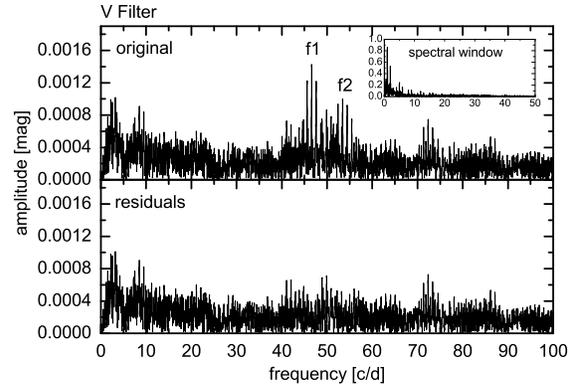}}
\caption{Amplitude spectrum of NGC 6530 5 in the $V$ filter; `f1' and `f2' mark the identified frequencies.}
\label{ngc6530_5amp}
\end{figure}

Two significant frequencies at 46.596\,c/d with amplitudes of 1.4\,mmag in $V$ and 1.8\,mmag in $B$ and at 53.417\,c/d with amplitudes of around 1\,mmag in both filters have been found and are clearly visible in the amplitude spectra (for the $V$ filter the amplitude spectra are shown in Fig. \ref{ngc6530_5amp}).

\subsection{NGC 6530 82}
NGC 6530 82 (WEBDA \#78) is located in field 1 of our observations, and again no spectral type or membership information is available in the literature. $UBV$ CCD photometry (Sung et al. \cite{sun00}) yields $V$\,=\,13.97\,mag, $(B-V)$\,=\,0.61\,mag, and $(U-B)$\,=\,0.30\,mag, which makes it a likely cluster member and places the star inside the classical instability strip in the HR diagram.

\begin{figure}[htb]
\centering{\includegraphics[width=8.5cm]{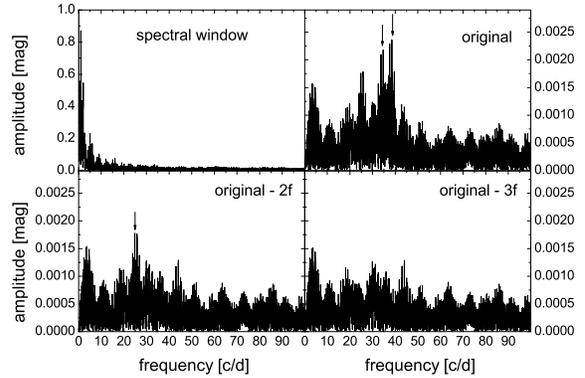}}
\caption{Amplitude spectrum of NGC 6530 82 in the $V$ filter; the identified and sequentially prewhitened frequencies are marked with arrows.}
\label{ngc6530_82amp}
\end{figure}

In the analysis, three frequencies between 24.829\,c/d and 38.531\,c/d with amplitudes between 1.7\,mmag and 2.8\,mmag have been found to be significant (see Table \ref{ngc6530-fa}). A simultaneous multi-sine fit with these three frequencies represents the shape of the light curve well. The frequencies are clearly detectable in both filters (the amplitude spectra of the $V$ filter are given in Fig. \ref{ngc6530_82amp}).

\subsection{NGC 6530 85}
NGC 6530 85 (WEBDA \#53) was also situated in field 1 of our observations and was studied by different authors.
Kilambi (1977) obtained $V$\,=\,12.96\,mag, $(B-V)$\,=\,0.62\,mag, $(U-B)$\,=\,0.30\,mag, and confirmed its cluster membership. These values match the $UBV$ CCD measurements of Sung et al. (\cite{sun00}), who found $V$\,=\,13.07\,mag, $(B-V)$\,=\,0.65\,mag, and $(U-B)$\,=\,0.35\,mag quite well.

The proper motion study by van Altena \& Jones (\cite{alt72}) yielded a membership probability of 78\% for NGC 6530 85 (i.e., their star \#173). They also reported an $E(B-V)$\,=\,0.35 mag and absolute magnitude $M_V$\,=\,+0.81\,mag. Chini \& Neckel (\cite{chi81}) determined a spectral type of A0 for NGC 6530 85.

\begin{figure}[htb]
\centering{\includegraphics[width=9.5cm]{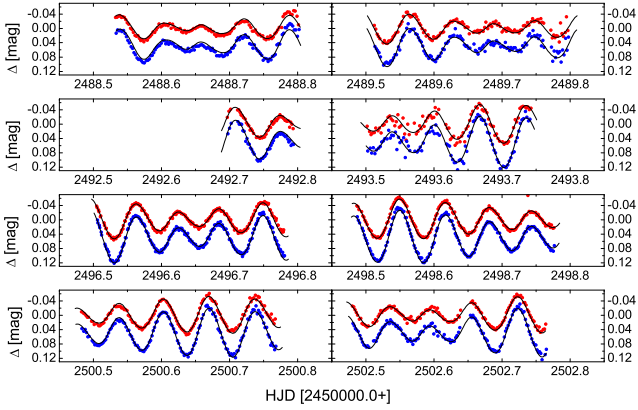}}
\caption{Differential light curve of NGC 6530 85 with a five frequency multi-sine fit (solid line); top: $V$ filter, bottom: $B$ filter (shifted for better visibility).}
\label{ngc6530_85lc}
\end{figure}

\begin{figure}[htb]
\centering{\includegraphics[width=8.5cm]{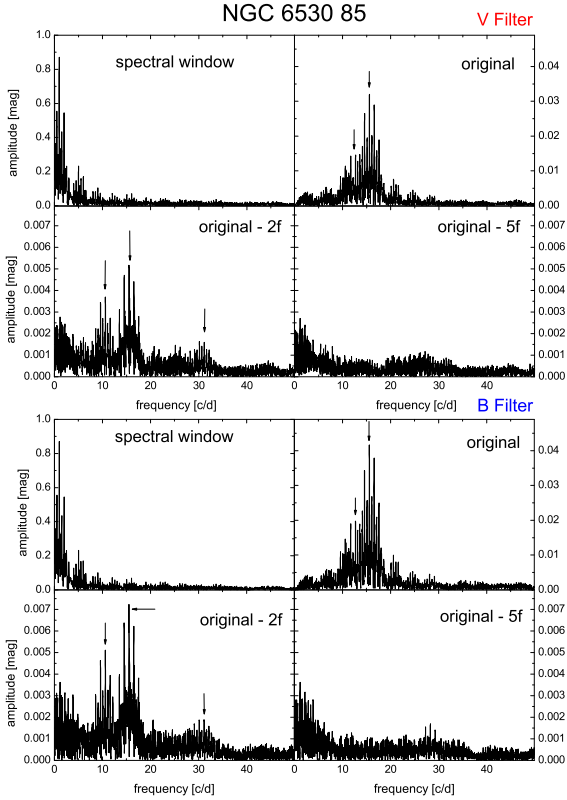}}
\caption{Amplitude spectra of NGC 6530 85 in both filters; the identified and sequentially prewhitened frequencies are marked with arrows.}
\label{ngc6530_85amp}
\end{figure}

The frequency analysis using {\sc Period98} and {\sc SigSpec} allowed us to identify simultaneous pulsation with five frequencies between 10.585\,c/d and 31.148\,c/d and amplitudes in the range of 39.1\,mmag to 1.8\,mmag (Table \ref{ngc6530-fa}, Fig. \ref{ngc6530_85amp}). The shape of the light curve demonstrates the multi-periodic nature of the star well (Fig. \ref{ngc6530_85lc}).

\subsection{NGC 6530 263}
NGC 6530 263 is one of the stars located in field 2 of our observations. $UBV$ CCD photometry (Sung et al. \cite{sun00}) yielded $V$\,=\,13.67\,mag, $(B-V)$\,=\,0.63\,mag, and $(U-B)$\,=\,0.42\,mag for star 263 (WEBDA \#57), which agree well with earlier measurements (e.g., from Kilambi \cite{kil77}). No information on the spectral type of the star is available in the literature.

NGC 6530 263 was also contained in the proper motion study by van Altena \& Jones (\cite{alt72}), but it belonged to the 135 stars that have not been included in their ``primary'' absolute parameter solution. They indicate that the star (their star \#178) might only be considered a {\it probable} cluster member.

\begin{figure}[htb]
\centering{\includegraphics[width=8.5cm]{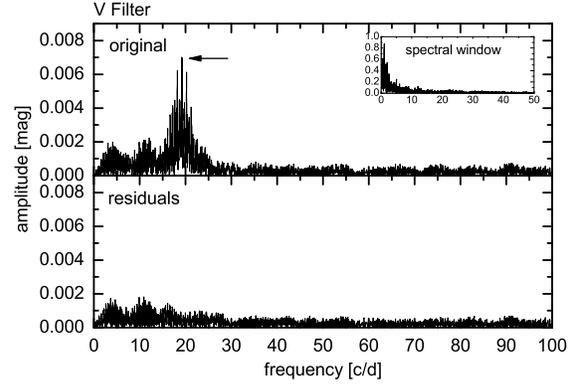}}
\caption{Amplitude spectrum of NGC 6530 263 in the $V$ filter; the identified frequency is marked with an arrow.}
\label{ngc6530_263amp}
\end{figure}

The frequency analysis of both filters showed a single significant frequency at 19.223\,c/d, corresponding to a period of 1.25 hours, with an amplitude of 8.3\,mmag in $B$ and 7.1\,mmag in $V$. This $\delta$ Scuti-like pulsation frequency clearly shows up in the amplitude spectra (e.g., for the $V$ filter see Fig. \ref{ngc6530_263amp}).

\subsection{NGC 6530 278}
NGC 6530 278 (WEBDA \#38) has also been one of the stars observed in field 2.
Kilambi (1977) obtained $V$\,=\,12.16\,mag, $(B-V)$\,=\,0.44\,mag, and $(U-B)$\,=\,0.39\,mag for this star, reported its cluster membership, and found indication for variability. The more recent values for $V$\,=\,12.17\,mag, $(B-V)$\,=\,0.53\,mag and $(U-B)$\,=\,0.36\,mag from CCD photometry (Sung et al. \cite{sun00}), confirm these values.

The proper motion study by van Altena \& Jones (\cite{alt72}) gave a membership probability of 68\% for NGC 6530 278 (their star \#157). They also report a value of \mbox{$E(B-V)$\,=\,0.35\,mag} and absolute magnitude $M_V$\,=\,-0.12\,mag. Chini \& Neckel (\cite{chi81}) used their $UBV$ and $H_{\beta}$ observations to derive a spectral type of A0 for this star.

\begin{figure}[htb]
\centering{\includegraphics[width=8.5cm]{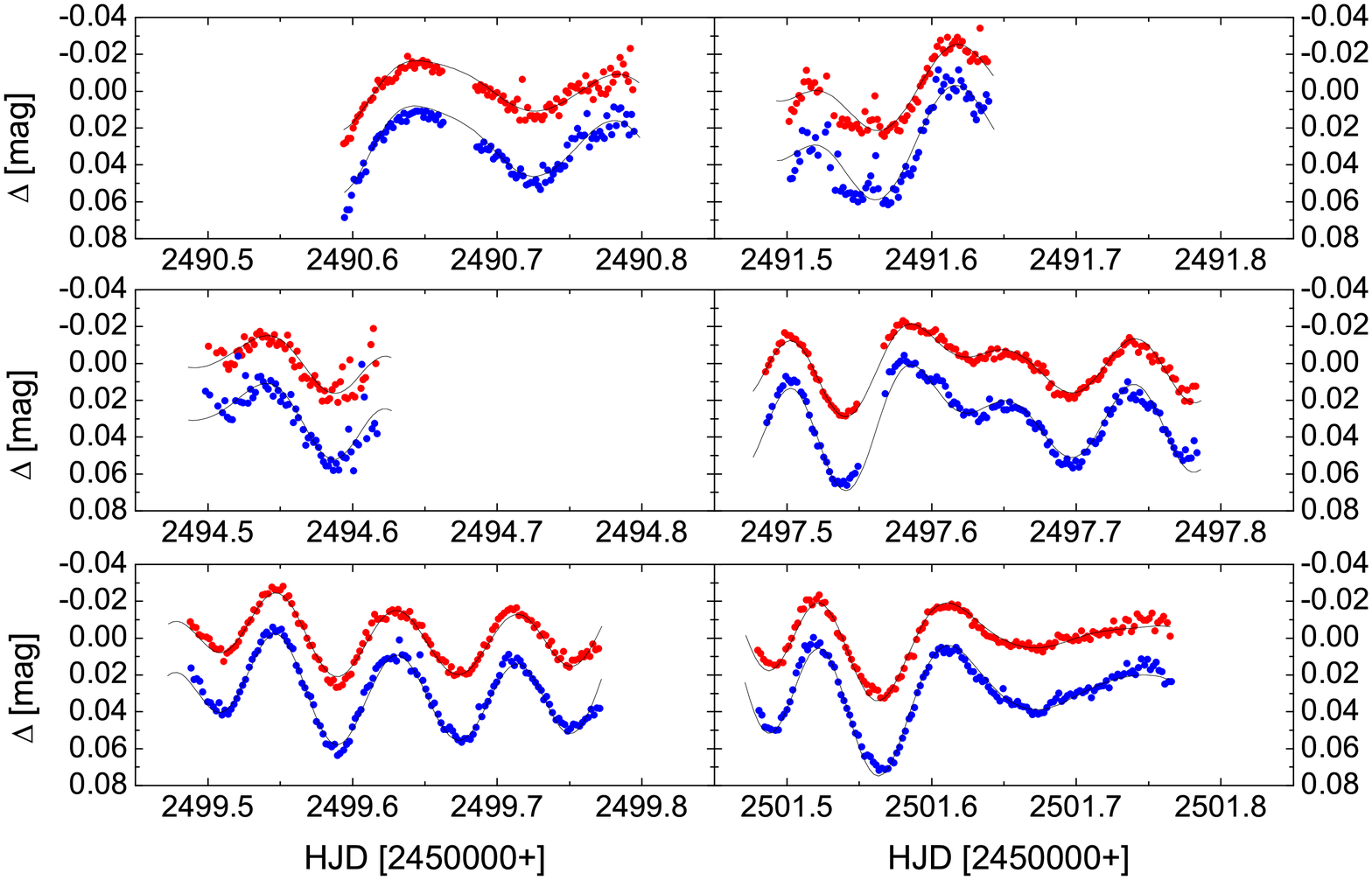}}
\caption{Differential light curve of NGC 6530 278 with a nine frequency multi-sine fit (solid line); top: $V$ filter, bottom: $B$ filter (shifted for better visibility).}
\label{ngc6530_278lc}
\end{figure}

\begin{figure}[htb]
\centering{\includegraphics[width=8.5cm]{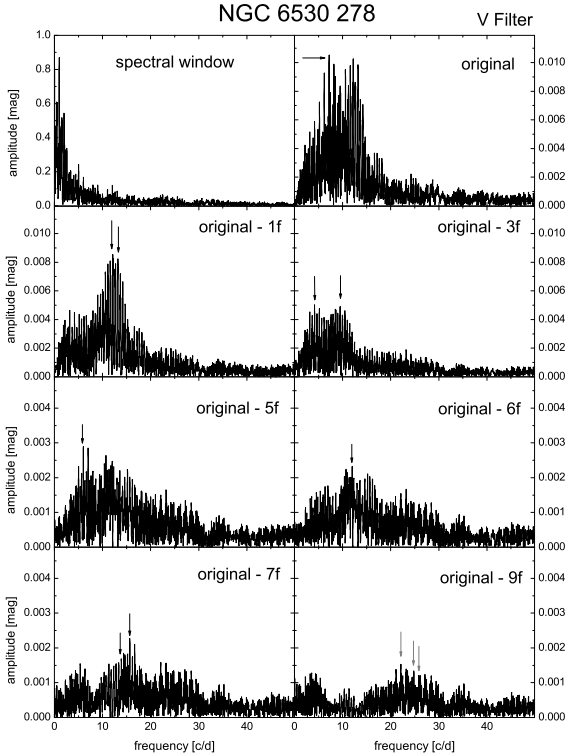}}
\caption{Amplitude spectra of NGC 6530 278 in the $V$ filter; the identified and sequentially prewhitened frequencies are marked with arrows.}
\label{ngc6530_278amp}
\end{figure}

The light curve of NGC 6530 278 (Fig. \ref{ngc6530_278lc}) clearly illustrates its multi-periodicity. The formal solution of the detailed frequency analysis yielded nine significant frequencies in both filters between 4.0\,c/d and 15.6\,c/d with amplitudes ranging from 13.8\,mmag down to 2.6\,mmag (see Table \ref{ngc6530-fa}, Fig. \ref{ngc6530_278amp}). The identification and prewhitening of intrinsic frequencies was difficult due to the numerous peaks caused by aliases in the frequency range between 0\,and 15\,c/d. At least three more frequencies between 20\,c/d and 30\,c/d might exist, but the variability of NGC 6530 278 is proven.

\subsection{NGC 6530 281}
NGC 6530 281 (WEBDA \#13) lies in field 2 of our observations. $UBV$ CCD photometry (Sung et al. \cite{sun00}) gives $V$\,=\,13.35\,mag, $(B-V)$\,=\,0.45\,mag, and $(U-B)$\,=\,0.27\,mag for this star, which agree well with values derived earlier by different authors. No spectral classification is available, but Kilambi (\cite{kil77}) reported that the star is most probably a cluster member. Its location in the cluster CMD and in the HR diagram confirm this suggestion.

\begin{figure}[htb]
\centering{\includegraphics[width=8.5cm]{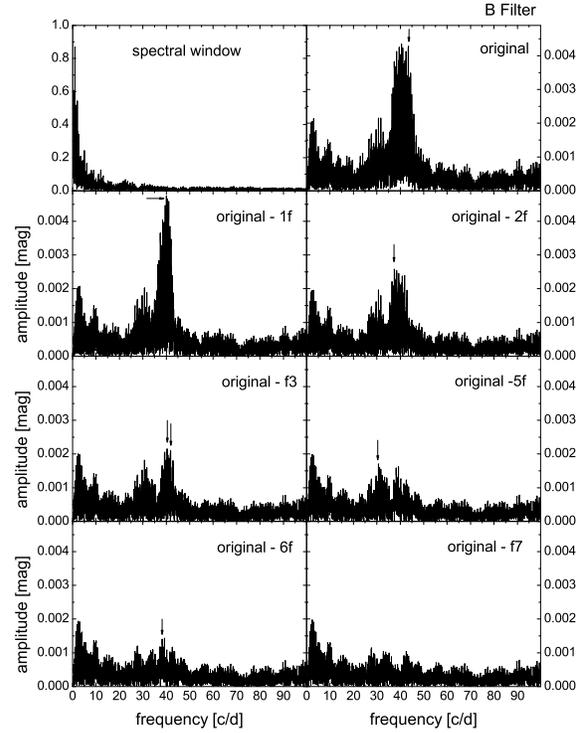}}
\caption{Amplitude spectrum of NGC 6530 281 in the $B$ filter; the identified and sequentially prewhitened frequencies are marked with arrows.}
\label{ngc6530_281B}
\end{figure}

Pulsation with seven periods was found as a formal solution. The frequencies lie between 30.6\,c/d and 43.4\,c/d, corresponding to periods of 47 to 33 minutes, where the amplitude spectrum in the $B$ filter is shown in Fig. \ref{ngc6530_281B}. The amplitudes range from 5.1\,mmag to 1.4\,mmag (see Table \ref{ngc6530-fa}). Similar to NGC 6530 278, the identification and prewhitening of intrinsic frequencies was difficult due to the numerous peaks caused by aliases in the frequency range between 35\,c/d and 45\,c/d.

\subsection{NGC 6530 288}
NGC 6530 288 is also located in field 2 of our observations.
With $V$\,= 13.23\,mag and $(B-V)$\,=\,0.41\,mag, NGC 6530 288 (WEBDA \#28) falls into the region of the classical instability strip in the HR diagram. Hence, it was also one of the prime candidates to search for PMS pulsation. In the $V$ filter data, a frequency of 17.996\,c/d corresponding to a period of 1.33 hours with an amplitude of 0.8\,mmag is significant. The same frequency can be found in the $B$ filter data, but it is not above 4$\cdot$S/N and has a {\sc SigSpec} significance of only 5.04. Longer time-series are needed to confirm the variability of this star.

\begin{table}[htb]
\caption{Frequencies and amplitudes determined for all PMS pulsators in NGC 6530, where the errors in the last digits of the corresponding quantities are given in parentheses.}
\label{ngc6530-fa}
\begin{center}
\begin{small}
\begin{tabular}{rcccc}
\hline
Star & No. & \multicolumn{1}{c}{Frequency} & \multicolumn{1}{c}{{\it V} amp.} & \multicolumn{1}{c}{{\it B} amp.}  \\
& & \multicolumn{1}{c}{[c/d]} & \multicolumn{1}{c}{[mmag]} & \multicolumn{1}{c}{[mmag]}  \\
 \hline
5  & f1 & 46.596(9) & 1.4(3) & 1.8(3)    \\
   & f2 & 53.417(9) & 1.0(3) & 1.1(3)   \\
\hline
82 & f1 & 38.531(6) & 2.4(3) & 2.8(3)   \\
   & f2 & 34.671(7) & 2.2(3) & 2.4(3)   \\
   & f3 & 24.829(9) & 1.8(3) & 1.7(3)   \\
\hline
85 & f1 & 15.579(1) & 30.2(3) & 39.1(3)    \\
   & f2 & 12.700(1) & 17.1(3) & 23.0(3)   \\
   & f3 & 15.531(2) &  8.2(3) & 11.4(3)    \\
   & f4 & 10.585(4) &  3.5(3) &  4.7(3)    \\
   & f5 & 31.148(8) &  1.8(3) &  2.0(3)   \\
\hline
263 & f1 & 19.223(2) & 7.1(2) & 8.3(3)  \\
\hline
278 & f1 & 7.200(2) & 6.6(2) & 9.4(3)  \\
    & f2 & 12.121(2) & 9.4(2) & 12.4(3)   \\
    & f3 & 13.218(2) & 9.9(2) & 13.8(3)    \\
    & f4 & 4.178(2) & 6.2(2) & 8.0(3)    \\
    & f5 & 9.488(3) & 5.0(2) & 6.6(3)    \\
    & f6 & 6.013(4) & 3.6(2) & 5.0(3)   \\
    & f7 & 11.984(3) & 4.9(2) & 6.7(3)    \\
    & f8 & 15.684(5) & 2.9(2) & 3.7(3)   \\
    & f9 & 13.896(6) & 2.6(2) & 3.5(3)   \\
\hline
281 & f1 & 43.418(4) & 4.2(2) & 4.7(2)   \\
    & f2 & 40.017(4) & 3.9(2) & 5.1(2)   \\
    & f3 & 37.457(6) & 2.4(2) & 2.7(2)   \\
    & f4 & 41.702(9) & 1.7(2) & 1.5(2)  \\
    & f5 & 40.367(8) & 1.9(2) & 2.2(2)   \\
    & f6 & 30.691(9) & 1.4(2) & 1.5(2)  \\
    & f7 & 38.245(7) & 2.1(2) & 2.1(2)   \\
\hline
\end{tabular}
\end{small}
\end{center}
\end{table}

\section{Pulsation constants}
The main objective of this work was to discover new members of the group of pulsating \pms stars, not to provide a complete asteroseismic investigation. Additional photometric time-series from multi-site campaigns or from space are needed to characterize the stars asteroseismologically in more detail.

\begin{table*}[htb]
\caption{Pulsation constant $Q$ for the newly discovered PMS pulsators (see text for detailed explanation).}
\label{qvalues}
\begin{center}
\begin{footnotesize}
\begin{tabular}{lccccccc}
\hline
Star & $M/$\Msun & $\log$ \Teff & $\log$ $L/$\Lsun & $\log g$ & $M_{bol}$ & Period & $Q$ \\
 &  &  &  &  & [mag] & [d] & [d] \\
\hline
IC 4996 37 & 2.26 & 3.892 & 1.17 & 4.136 & 1.820 & 0.031 & 0.015  \\
IC 4996 40 & 2.50 & 3.904 & 1.31 & 4.088 & 1.474 & 0.030 & 0.013  \\
NGC 6530 5 & 2.31 & 3.923 & 1.20 & 4.239 & 1.747 & 0.022 & 0.012  \\
 &  &  &  &  &  & 0.019 & 0.011  \\
NGC 6530 82 & 2.00 & 3.875 & 1.01 & 4.170 & 2.215 & 0.026 & 0.014  \\
 &  &  &  &  &  & 0.029 & 0.015  \\
 &  &  &  &  &  & 0.040 & 0.021  \\
NGC 6530 85 & 2.62 & 3.864 & 1.37 & 3.887 & 1.326 & 0.064 & 0.020  \\
 &  &  &  &  &  & 0.079 & 0.024  \\
 &  &  &  &  &  & 0.064 & 0.020  \\
 &  &  &  &  &  & 0.095 & 0.029 \\
 &  &  &  &  &  & 0.032 & 0.010  \\
NGC 6530 263 & 2.19 & 3.869 & 1.13 & 4.069 & 1.921 & 0.052 & 0.023  \\
NGC 6530 278 & 3.47 & 3.896 & 1.75 & 3.762 & 0.383 & 0.139 & 0.032  \\
 &  &  &  &  &  & 0.083 & 0.019  \\
 &  &  &  &  &  & 0.076 & 0.017  \\
 &  &  &  &  &  & 0.239 & 0.055  \\
 &  &  &  &  &  & 0.105 & 0.024  \\
 &  &  &  &  &  & 0.166 & 0.038  \\
 &  &  &  &  &  & 0.083 & 0.019  \\
 &  &  &  &  &  & 0.064 & 0.015  \\
 &  &  &  &  &  & 0.072 & 0.017  \\
NGC 6530 281 & 2.47 & 3.918 & 1.29 & 4.156 & 1.519 & 0.023 & 0.011  \\
 &  &  &  &  &  & 0.025 & 0.012  \\
 &  &  &  &  &  & 0.027 & 0.013  \\
 &  &  &  &  &  & 0.024 & 0.012  \\
 &  &  &  &  &  & 0.025 & 0.012  \\
 &  &  &  &  &  & 0.033 & 0.016 \\
 &  &  &  &  &  & 0.026 & 0.013  \\
\hline
\end{tabular}
\end{footnotesize}
\end{center}
\end{table*}

Nevertheless, for classical \dsct stars, the pulsation constant, $Q$, can be used to distinguish radial fundamental or higher overtone modes. As shown by Breger (\cite{bre79}), $Q$ can be computed using the following relation:

\begin{equation}
{\rm log}\,Q = -6.454 + {\rm log}\,P + 0.5\,{\rm log}\,g + 0.1\,M_{bol} + {\rm log}\,T_{\rm eff}.
\end{equation}

Using the empirical calibration by Reed (\cite{ree98}), which is described below, it is possible to derive $M_{bol}$, $\log$ \Teff, and \mbox{log L/\Lsun} \,\,from \bv0 and $M_v$ for all the newly discovered pulsating \pms stars. For stars with $\log$ \Teff $\le$ 3.961 (i.e., \Teff $\le$ 9140 K) the relation
\begin{equation}
(B-V)_0 = -3.648\,(\log T_{\rm eff}) + 14.551
\end{equation}
can be used to determine $\log T_{\rm eff}$. The bolometric magnitude, $M_{bol}$, is further defined as:
\begin{equation}
M_{bol} = M_v + BC(T) ,
\end{equation}
where the bolometric correction, $BC(T)$, can be determined from
\begin{small}
\begin{eqnarray}
BC(T) = -8.499\,(\log T_{\rm eff}-4)^4 + 13.421\,(\log T_{\rm eff}-4)^3 -  \nonumber\\
- 8.131\,(\log T_{\rm eff}-4)^2 - 3.901\,(\log T_{\rm eff}-4) - 0.438 .
\end{eqnarray}
\end{small}
To compute the pulsation constant, $Q$, the mass ratios and $\log g$ values are needed. Voigt (\cite{voi91}), e.g., gives:
\begin{equation}
{\rm log} (M/M_{\odot}) = 0.59 - 0.13\,M_{bol}
\end{equation}
\begin{equation}
g/g_{\odot} = (M/M_{\odot})\cdot(T_{\rm eff}^4 / T_{\rm eff,\odot}^4) / (L/L_{\odot}),
\end{equation}
with $T_{\rm eff,\odot}$ = 5780 K and $g_{\odot} = 2.7 \cdot 10^4 \, {\rm cms}^{\rm {-1}}$.

The radial $Q$ values are 0.033\,d for the fundamental mode, 0.025\,d for the first, 0.020\,d for the second, and 0.017\,d for the third overtones; values smaller than 0.017\,d are identified as higher overtones. Regarding the resulting $Q$ values given in Table \ref{qvalues}, it becomes evident that most of the detected periods correspond to higher overtone modes.
Only the first period of NGC 6530 278 (P\,=\,0.139\,d) and possibly the fourth period of NGC 6530 85 (P\,=\,0.094\,d) seem to be radial fundamental modes.

\section{Other variable stars}
Several other cluster members show variability on very different timescales. Some other objects are variable, but may not be cluster members, or their variability remains inconclusive in the data. For completeness, the most interesting other variable stars, which are not $\delta$ Scuti like PMS pulsators, are discussed below in detail. All definitive and suspected variables are listed in the corresponding tables (Tables \ref{ic4996var} and \ref{ngc6530var}).

\subsection{Other variable stars in IC 4996}

\subsubsection{IC 4996 67}
\begin{figure}[htb]
\centering
\includegraphics[width=8.5cm]{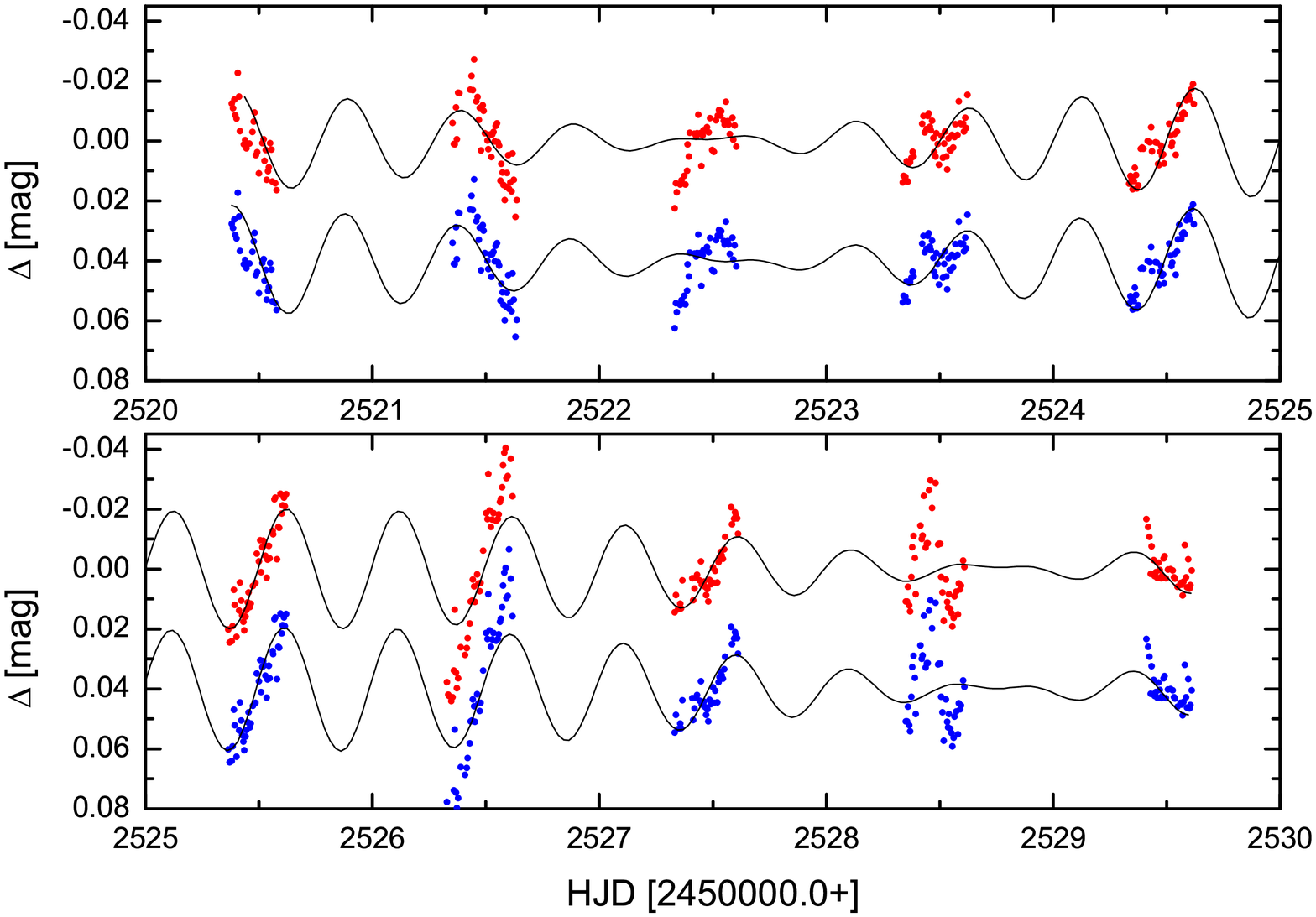}
\caption{Differential light curve of IC 4996 67 in $V$ (top) and $B$ filters (bottom, shifted for better visibility). The solid line corresponds to the multi-sine fit with the two significant frequencies.}
\label{ic4996-67lc}
\end{figure}

IC 4996 67 has $V$\,=\,12.82\,mag and $(B-V)$\,=\,0.52\,mag (Delgado et al. \cite{del98}), but no spectral classification exists for this star. The light curves in both filters look variable with relatively high amplitudes (Fig. \ref{ic4996-67lc}). The frequency analyses using {\sc Period98} and {\sc SigSpec} yielded two significant frequencies, which seem to be intrinsic. The first frequency is at 1.928\,c/d with amplitudes of 10.3\,mmag in $V$ and 10.5\,mmag in $B$, and the second at 2.088\,c/d with amplitudes of 9.7\,mmag in $V$ and 10.2\,mmag in $B$. Permitting the star to be less distant than the cluster, it could be a foreground \gdor type star. Further studies are needed to determine the nature of its variability, and a reliable spectral classification would be most important.

\subsubsection{IC 4996 80}
\begin{figure}[htb]
\centering
\includegraphics[width=8cm]{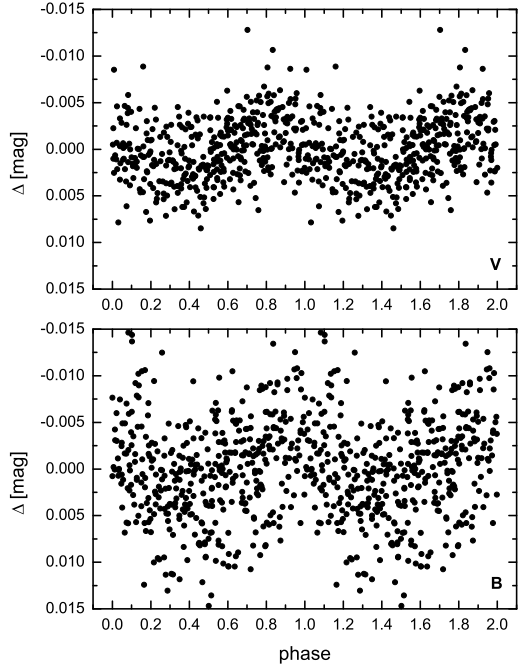}
\caption{Phase plots for IC 4996 80 in $V$ (top) and $B$ (bottom) filters with a frequency of 3.578 c/d.}
\label{ic4996-80pha}
\end{figure}

IC 4996 80 is located at the blue edge of the instability region in the HR diagram. $V$\,=\,14.04\,mag, $(B-V)$\,=\,0.58\,mag, and $(U-B)$\,=\,0.15\,mag were derived by Delgado et al. (\cite{del98}), but again no spectral classification is available. In the data sets of both filters, one significant frequency at 3.578\,c/d, i.e., a period of $\sim$6.7 hours, with an amplitude of 2.0\,mmag in $V$ and 2.9\,mmag in $B$ was detected (Fig. \ref{ic4996-80pha}).

If star 80 is slightly hotter than indicated from its color, it could fall into the region of the slowly pulsating B star (SPB-) instability domain in the HR diagram.

\subsubsection{IC 4996 86}
IC 4996 86 is again bluer than the blue edge of the classical instability region and has $V$\,=\,13.87\,mag, $(B-V)$\,=\,0.47\,mag, and $(U-B)$\,=\,0.08\,mag (Delgado et al. \cite{del98}). Two frequencies at 3.294\,c/d (i.e., a period of 7.3 hours) and 3.380\,c/d (i.e., a period of 7.1 hours) indicate a possible $\beta$ Cephei- or SPB-type variability.

\subsubsection{IC 4996 95}
\begin{figure}[htb]
\vspace{10mm}
\centering
\includegraphics[width=8.5cm]{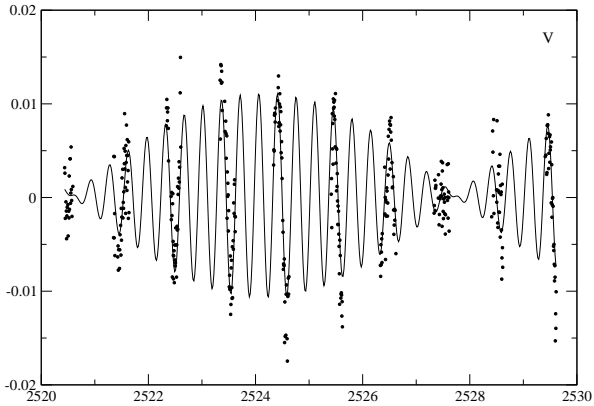}
\caption{Differential light curve of IC 4996 95 in $V$ with a multi-sine fit (solid line) of the two significant frequencies.}
\label{ic4996-95lc}
\end{figure}

For IC 4996 95, no magnitudes and colors are available from the literature. Our calibration yields $V$\,=\,13.72\,mag and $(B-V)$\,=\,0.61\,mag, which puts the star outside the classical instability strip. The two significant frequencies at 2.807\,c/d (with 5.4\,mmag amplitude in $V$ and $B$) and 2.942\,c/d (with 6.9\,mmag amplitude in $V$ and 5.1\,mmag in $B$) lie very close to each other and produce a beat in the light curve (Fig. \ref{ic4996-95lc}). The corresponding periods are 8.6 and 8.2 hours, respectively. It also seems that the star is not a cluster member, and the origin of its variation can only be suspected. It may be caused by $\beta$ Cephei- or SPB-type pulsations.

\subsubsection{IC 4996 104}
\begin{figure}[htp]
\centering
\includegraphics[width=8.5cm]{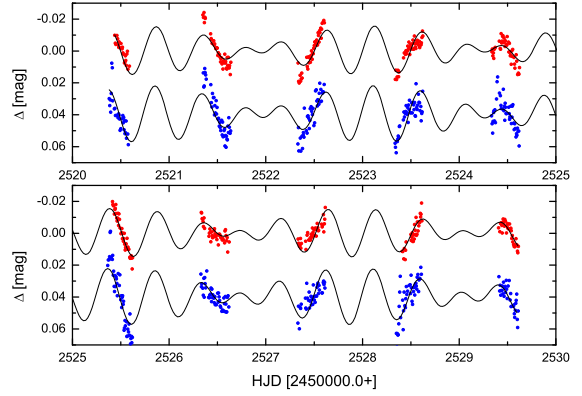}
\caption{Differential light curves for IC 4996 104 in $V$ (top) and $B$ (bottom, shifted for better visibility) filters; the solid line corresponds to a multi-sine fit with the two significant frequencies.}
\label{ic4996-104lc}
\end{figure}

With $V$\,=\,14.03\,mag, $(B-V)$\,=\,0.64\,mag, and $(U-B)$\,=\,0.02\,mag the star is also near the blue edge of the classical instability region. Two frequencies at 2.209\,c/d (i.e., a period of 10.9 hours) with an amplitude of 7.5\,mmag in $V$ and 10.4\,mmag in $B$, and at 1.786\,c/d (i.e., a period of 13.4 hours) with an amplitude of 6.6\,mmag in $V$ and 7.7\,mmag in $B$ were found to be significant, and a simultaneous multi-sine fit to the data reproduces the shape of the light curve well (Fig. \ref{ic4996-104lc}). There is no spectral class available, but it could be a SPB star.

\begin{table}[htb]
\caption{Other variables and suspected variables in the field of IC 4996: `Star' denotes our star number and `Ref' the cross-reference with the literature (D ... Delgado et al. \cite{del98}, W ... number given in the WEBDA database). Values marked with asterisks were derived from our calibration, when no values were available in the literature.}
\label{ic4996var}
\begin{scriptsize}
\begin{center}
\begin{tabular}{rllccl}
\hline
 Star & \multicolumn{1}{c}{Ref} & \multicolumn{1}{c}{$V$} & \multicolumn{1}{c}{$(B-V)$} & var. in filter & Remarks \\
 \# & & \multicolumn{1}{c}{mag} & \multicolumn{1}{c}{mag} & $B$ / $V$ / $B$ \& $V$ & \\
 \hline
 10 & D 73 & 14.62 & 0.84 & $V$ & inconclusive \\
 18 & D 54 & 12.36 & 0.66 & $B$ \& $V$ & different $P$ in $B$ \& $V$ \\
 19 & D 56 & 12.93 & 0.39 & $B$ \& $V$ & different $P$ in $B$ \& $V$ \\
 38 & D 91 & 15.63 & 1.86 & $B$ & inconclusive \\
 42 & D 13 & 13.34 & 0.52 & $B$ \& $V$ & different $P$ in $B$ \& $V$ \\
 51 & D 14 & 13.66 & 0.59 & $B$ & inconclusive \\
 67 & D 55 & 12.82 & 0.52 & $B$ \& $V$ & 2 periods \\
 74 & D 83 & 15.25 & 1.13 & $V$ & $P$ is not significant in $B$ \\
 80 & D 18 & 14.04 & 0.58 & $B$ \& $V$ & $P$ $\sim$ 6.708 h \\
 81 & D 61 & 13.60 & 0.59 & $B$ \& $V$ & inconclusive \\
 86 & D 17 & 13.87 & 0.47 & $V$ & 2 periods, unclear\\
 89 & W 110 & 13.81 & 0.55 & $V$ & $P$ is not significant in $B$ \\
 95 & W 50 & 13.72$^{\ast}$ & 0.61$^{\ast}$ & $B$ \& $V$ & 2 close periods\\
 104 & W 115 & 14.03 & 0.64 & $B$ \& $V$ & 2 periods \\
 108 & D 96 & 15.71 & 0.89 & $V$ & inconclusive \\
\hline
\end{tabular}
\end{center}
\end{scriptsize}
\end{table}

\subsection{Other variable stars in NGC 6530}

\subsubsection{NGC 6530 13}
\begin{figure}[htb]
\centering
\includegraphics[width=8.5cm]{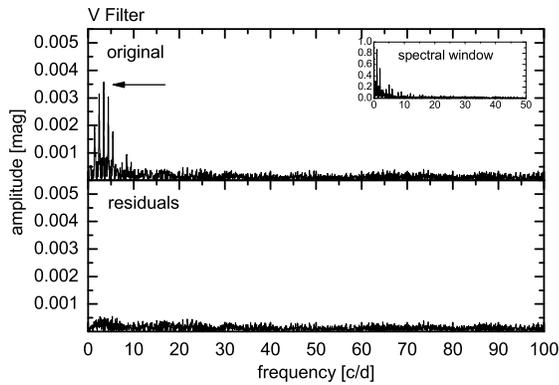}
\caption{Amplitude spectrum of NGC 6530 13 in the $V$ filter, where the significant frequency is indicated with an arrow.}
\label{ngc6530-13amp}
\end{figure}

NGC 6530 13 ($V$\,=\,12.65\,mag) was located in the overlapping region of the CCDs. It was classified as a B8 star by van den Ancker et al. (\cite{van97}). Van Altena \& Jones (\cite{alt72}) estimate its cluster membership probability to be only 25\%.
The frequency analysis yielded a single significant frequency at 3.432\,c/d, corresponding to a period of 7.0 hours, with amplitudes of 3.6\,mmag in $V$ (Fig. \ref{ngc6530-13amp}) and 4.7\,mmag in $B$. The star could be a SPB-type variable.

\subsubsection{NGC 6530 21}
\begin{figure}[htb]
\centering
\includegraphics[width=8.5cm]{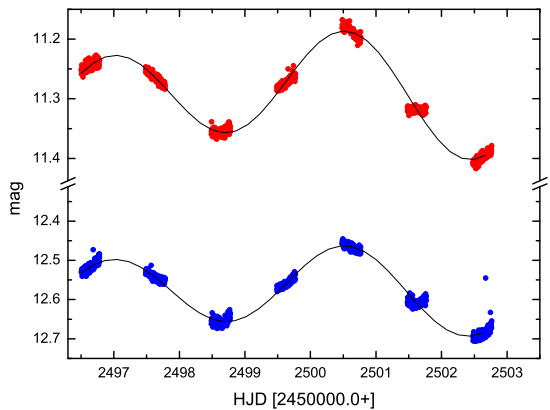}
\caption{Part of the light curve of NGC 6530 21 in the $V$ (top) and $B$ filters (bottom).}
\label{ngc6530-21lc}
\end{figure}

NGC 6530 21 was observed in the overlapping field of the CCDs. With $V$\,=\,14.17\,mag and $(B-V)$\,=\,0.97\,mag, it is located redwards of the instability region in the HR diagram. No spectral type or membership information is available for this star, but from its $(B-V)_0$\,=\,0.62\,mag (dereddened with $E(B-V)$\,=\,0.35\,mag, adopted from Sung et al. \cite{sun00}) it should be an early G spectral type with log \Teff = 3.77. Fourier analysis with {\sc Period98} yields two significant frequencies in the data sets of both filters, which are confirmed by {\sc SigSpec}. A simultaneous multi-sine fit to the data with the two significant frequencies detected with both methods, at 0.277\,c/d (i.e., a period of 3.613 days) with amplitudes of 8.7\,mmag in $V$ and 9.7\,mmag in $B$ and at 0.166\, c/d (i.e., a period of 6.037 days) with amplitudes of 2.8\,mmag in $V$ and 2.3\,mmag in $B$ reproduces the shape of the light curve well (Fig. \ref{ngc6530-21lc}).

\subsubsection{NGC 6530 55}
\begin{figure}[htb]
\centering
\includegraphics[width=8.5cm]{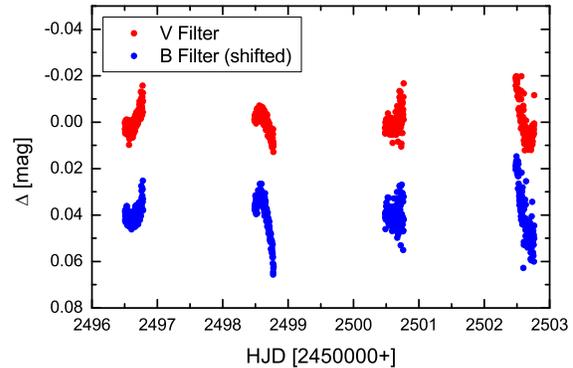}
\caption{Part of the differential light curve of NGC 6530 55 in the $V$ (top) and $B$ filters (bottom, shifted for better visibility).}
\label{ngc6530-55lc}
\end{figure}

NGC 6530 55 is situated bluewards of the instability region and was classified as spectral type B6 by van den Ancker et al. (\cite{van97}). The shape of the light curve clearly indicates variability (see Fig. \ref{ngc6530-55lc}). The frequency analyses using {\sc Period98} and {\sc SigSpec} detect a single significant frequency of 2.285\,c/d corresponding to a period of 10.5 hours.
Such a variability could be explained by SPB type pulsation.

\subsubsection{NGC 6530 73}
\begin{figure}[htb]
\centering
\includegraphics[width=8.5cm]{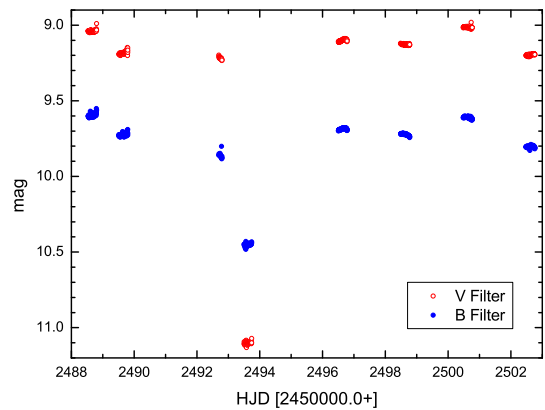}
\caption{Light curve of the suspected eclipsing binary NGC 6530 73 in the $V$ (open circles) and $B$ (solid circles) filters.}
\label{ngc6530-73}
\end{figure}

NGC 6530 73 ($V$\,=\,11.94\,mag) was classified as spectral type B5e by van den Ancker et al. (\cite{van97}). From the shape of its light curve, it is most likely a binary star: as the eclipse is deeper in the $V$ than in the $B$ filter, the redder star seems to occult its bluer companion (see Fig. \ref{ngc6530-73}).

\begin{table}[htb]
\caption{Other variables and suspected variables in the field of NGC 6530: `Star' denotes our star number and `Ref' the cross reference with numbers given in the WEBDA database, $V$ and $(B-V)$ values were taken from the literature. }
\label{ngc6530var}
\begin{scriptsize}
\begin{center}
\begin{tabular}{rlcccl}
\hline
 Star & \multicolumn{1}{c}{Ref} & $V$ & $(B-V)$ & var. in filter & remarks \\
 \# & \multicolumn{1}{c}{WEBDA} & mag & mag & $B$ / $V$ / $B$ \& $V$ & \\
 \hline
 13 & W 94 & 12.65 & 0.27 & $B$ \& $V$ & late B type star \\
 21 & W 75 & 14.17 & 0.97 & $B$ \& $V$ & 2 periods\\
 42 & W 41 & 12.53 & 0.34 & $B$ \& $V$ & inconclusive\\
 55 & W 84 & 11.87 & 0.22 & $B$ \& $V$ & 1 (4) periods ?\\
 70 & W 1681 & 15.08 & 1.30 & $B$ \& $V$ & binary ?\\
 73 & W 114 & 11.94 & 0.39 & $B$ \& $V$ & binary\\
 81 & W 79 & 14.59 & 1.18 & $B$ \& $V$ & binary ?\\
 83 & W 1641 & 13.77 & 1.09 & $B$ \& $V$ & 1 $P$ \\
 88 & W 1504 & 14.26 & 1.22 & $B$ \& $V$ & 2 periods \\
 93 & W 1458 & 14.61 & 1.08 & $B$ \& $V$ & different $P$ in $B$ \& $V$\\
 96 & W 40 & 13.45 & 1.20 & $B$ \& $V$ & 1 $P$, inconclusive\\
 109 & W 112 & 14.84 & 1.02 & $B$ \& $V$ & different $P$ in $B$ \& $V$\\
 117 & W 141 & 11.60 & 0.45 & $B$ \& $V$ & different $P$ in $B$ \& $V$\\
 119 & W 116 & 11.30 & 0.41 & $B$ \& $V$ & 1 $P$, inconclusive\\
 248 & W 19 & 10.76 & 0.24 & $B$ \& $V$ & suspected \\
 252 & W 165 & 14.61 & 1.06 & $B$ \& $V$ & 1 $P$, inconclusive\\
 256 & W 20 & 14.10 & 0.70 & $B$ \& $V$ & suspected \\
 274 & W 144 & 11.20 & 0.21 & $B$ \& $V$ & 2 periods \\
 285 & W 230 & 14.09 & 0.71 & $B$ \& $V$ & inconclusive \\
 305 & W 151 & 12.19 & 0.72 & $B$ \& $V$ & different $P$ in $B$ \& $V$\\
\hline
\end{tabular}
\end{center}
\end{scriptsize}
\end{table}

\section{Cluster properties}
\subsection{IC 4996}
\begin{figure}[htb]
\centering
\includegraphics[width=8.5cm]{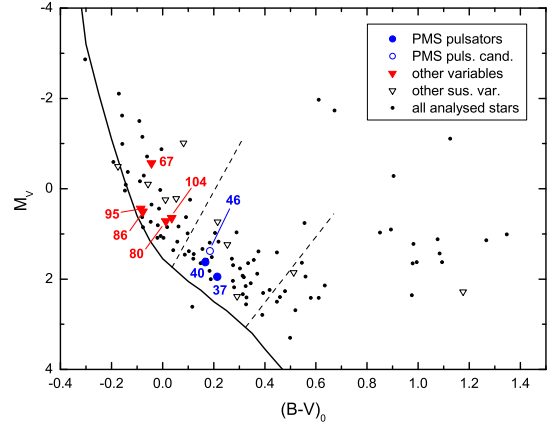}
\caption{Observational HR diagram of all stars in the field of IC 4996: bona fide PMS pulsators (filled circles), candidate PMS pulsator (open circle), other variables (filled triangles), and other suspected variables (open triangles); the ZAMS values are taken from Schmidt-Kaler (solid line), and the borders of the classical instability strip (dashed lines) have been transformed into the $M_V-(B-V)_0$ plane.}
\label{ic4996-hrd-vars}
\end{figure}

40 stars lie within the region of the classical instability strip in IC 4996, and hence have been prime candidates to search for PMS pulsation. Two stars, IC 4996 37 and IC 4996 40, had large enough amplitudes to allow for an unambiguous detection of pulsation, while for IC 4996 46 pulsation can only be suspected. From the location of the PMS pulsators in the HR diagram, they are most likely members of the cluster (Fig. \ref{ic4996-hrd-vars}). A definite decision on their cluster membership can only be drawn after a detailed proper motion or radial velocity study.
However, the fraction of pulsating \pms stars corresponds to $\sim$10\%.
As the observations have been performed in service mode and the integration time per frame was not long enough for stars in the magnitude range between 14 $< V <$ 16 mag, additional PMS pulsators may be discovered in a follow-up observing run for data with a lower noise level.

Of the total 113 stars analyzed in the field of the cluster, 16 stars have been found to be variable, including the PMS pulsators, which means only $\sim$15\%. The reason for such a low percentage lies in the higher noise level of the measurements: Especially towards the fainter, and hence redder stars, the noise level increased dramatically and made it impossible to detect variability.
Figure \ref{ic4996-hrd-vars} shows the location of all detected variables and suspected variable stars in IC 4996, where the lack of variable, red stars is clearly illustrated.

\begin{figure}[htb]
\centering
\includegraphics[width=8.5cm]{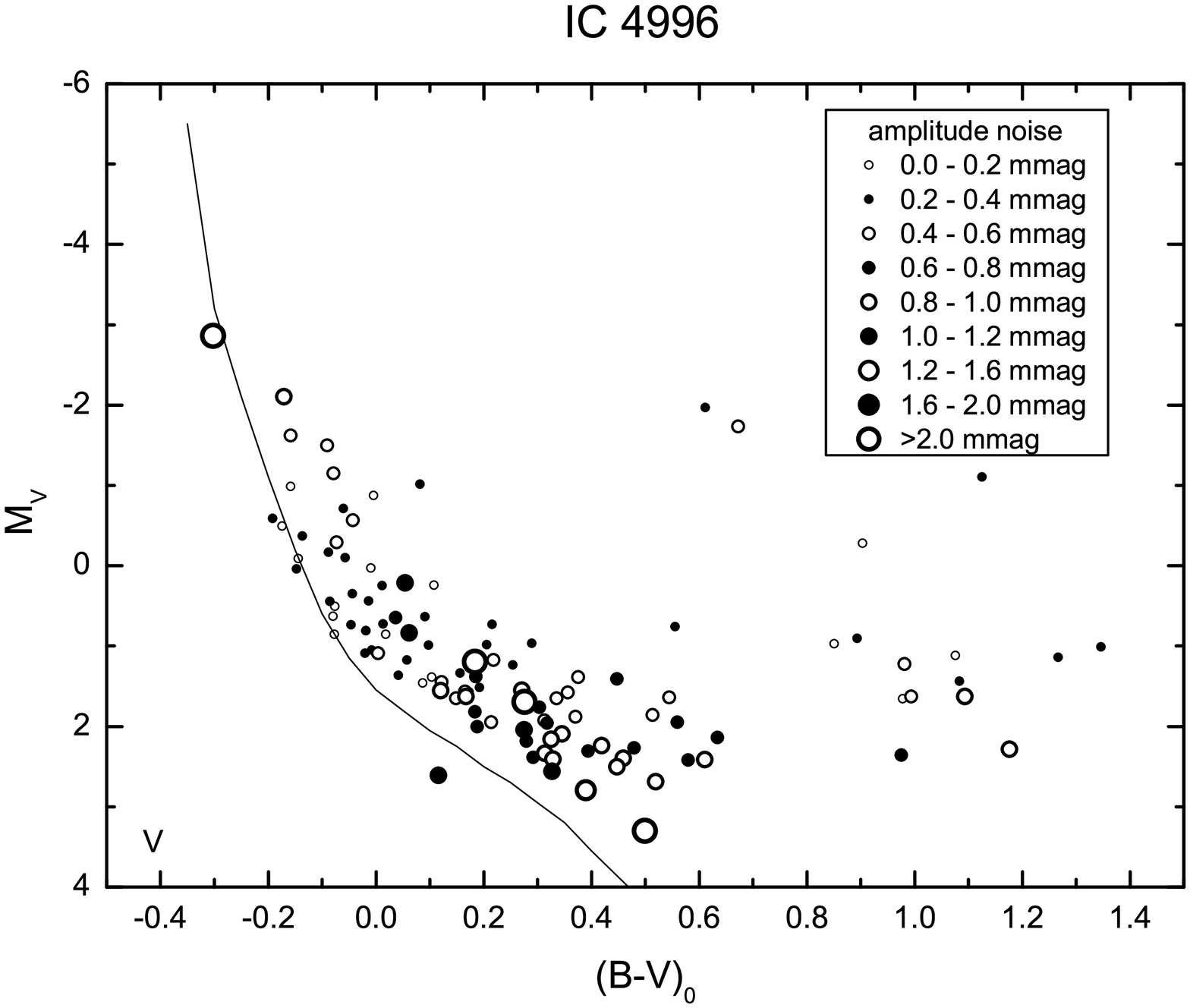}
\includegraphics[width=8.5cm]{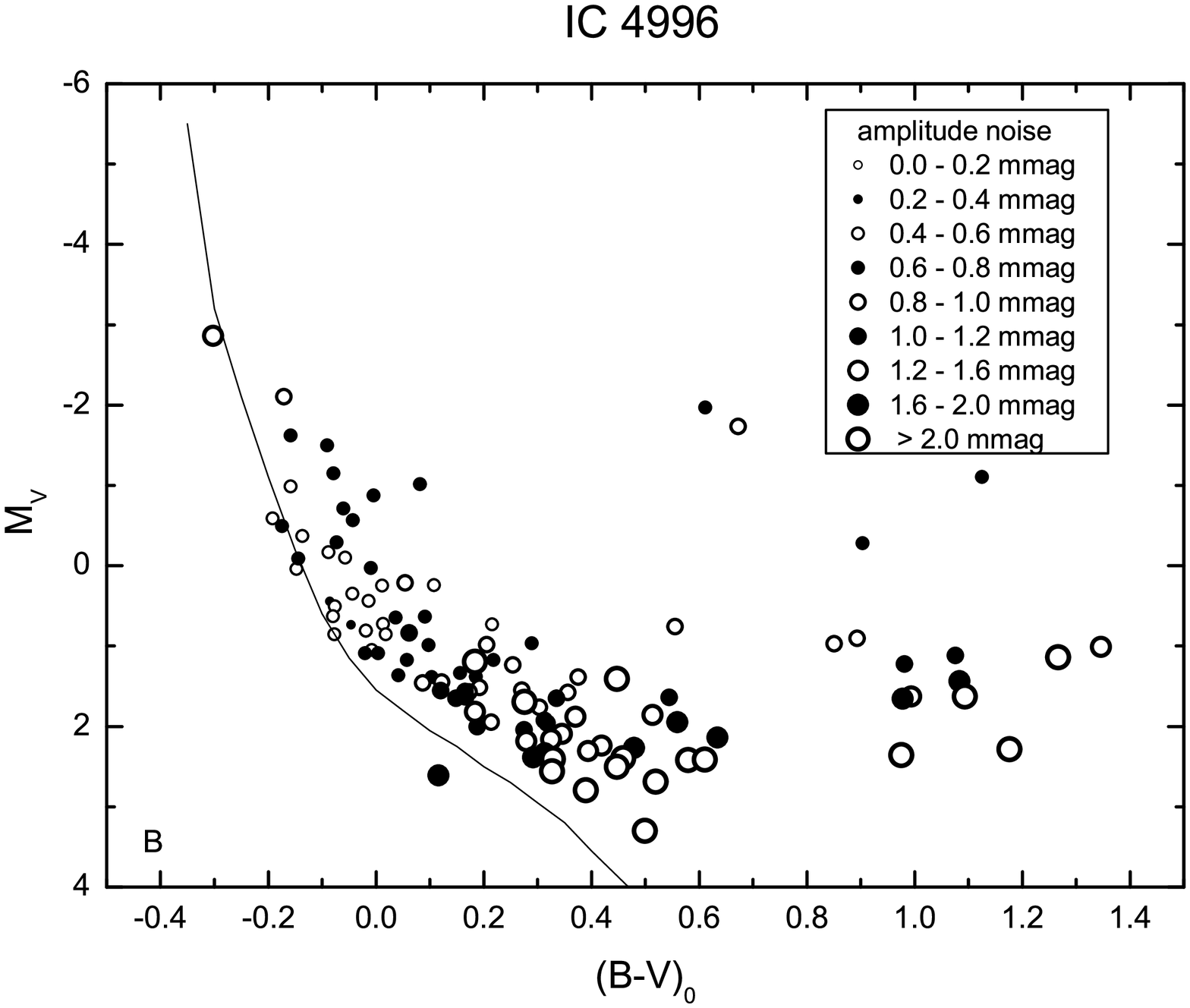}
\caption{HR diagram of all measured stars in the field of IC 4996, where the stars have different symbols and sizes according to their expected amplitude noise levels; {\it top:} $V$ filter, {\it bottom:} $B$ filter; the values for the ZAMS (solid line) are taken from Schmidt-Kaler (\cite{sch65}).}
\label{ic4996-hrd-amp}
\end{figure}

Figure \ref{ic4996-hrd-amp} shows HR diagrams for IC 4996 (top: $V$ and bottom: $B$ filters), where the stars have different symbols and sizes according to their amplitude noise levels ($amp_{noise}$) based on their time-domain point-to-point scatter ($\sigma_{pt2pt}$). The amplitude noise level in the Fourier domain is given as:

\begin{equation}
amp_{noise} = 2 \cdot \sigma_{pt2pt} / \sqrt{N}
\end{equation}

where $N$ denotes the number of data points per star.
The brightest star, IC 4996 34, has a $V$\,=\,10.45\,mag and was partly saturating the CCD chip, which is the reason why it has one of the highest values for the amplitude noise.

\subsection{NGC 6530}
In NGC 6530, six bona fide $\delta$ Scuti-like pulsating \pms stars could be discovered. Of those pulsators, one star is located in the overlapping region, two in field 1, and three in field 2, which justified the decision to split the observing time for two different fields. For NGC 6530 288, which is situated in field 2 of the observations, the data quality is not sufficient to unambiguously decide on its pulsation. Hence, it is only a likely candidate for a pulsating PMS star.
Taking into account that 25 stars have been primary candidates for pulsation because they are located in the region of the classical instability strip, the six pulsating stars correspond to 24\%. This percentage is higher than for the other clusters we analyzed. It is also closer to the fractions of 1/3 to 1/2 pulsating stars in the lower instability strip reported by Breger (\cite{bre00}) for the classical, post-, and main-sequence \dsct stars.
Whether this discrepancy is of instrumental origin or has an astrophysical background connected with the evolution of such young stars, cannot yet be decided.

\begin{figure}[htb]
\centering
\includegraphics[width=8.5cm]{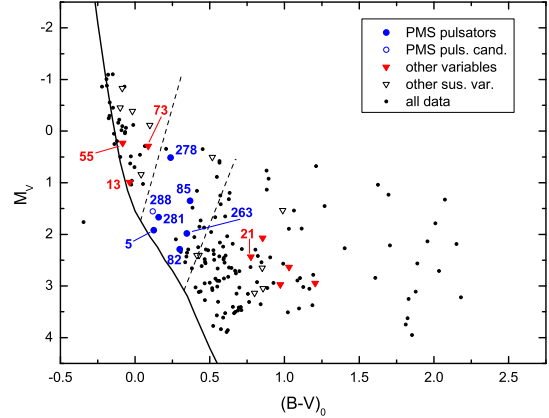}
\caption{Observational HR diagram of all stars in the field of NGC 6530: bona fide PMS pulsators (filled circles), candidate PMS pulsator (open circle), other variables (filled triangles), and other suspected variables (open triangles); the ZAMS values are taken from Schmidt-Kaler (solid line) and the borders of the classical instability strip (dashed lines) have been transformed into the $M_V-(B-V)_0$ plane.}
\label{ngc6530-hrd-vars}
\end{figure}

Among the 194 observed stars in the fields of NGC 6530, at least 12 other stars were found to be variable.
Figure \ref{ngc6530-hrd-vars} shows the location of all detected variable and suspected variable stars in NGC 6530 in the HR diagram, and it becomes evident that some stars in the field of the cluster might not be members, but are fore- or background field stars.

\begin{figure}[htb]
\centering
\includegraphics[width=8.5cm]{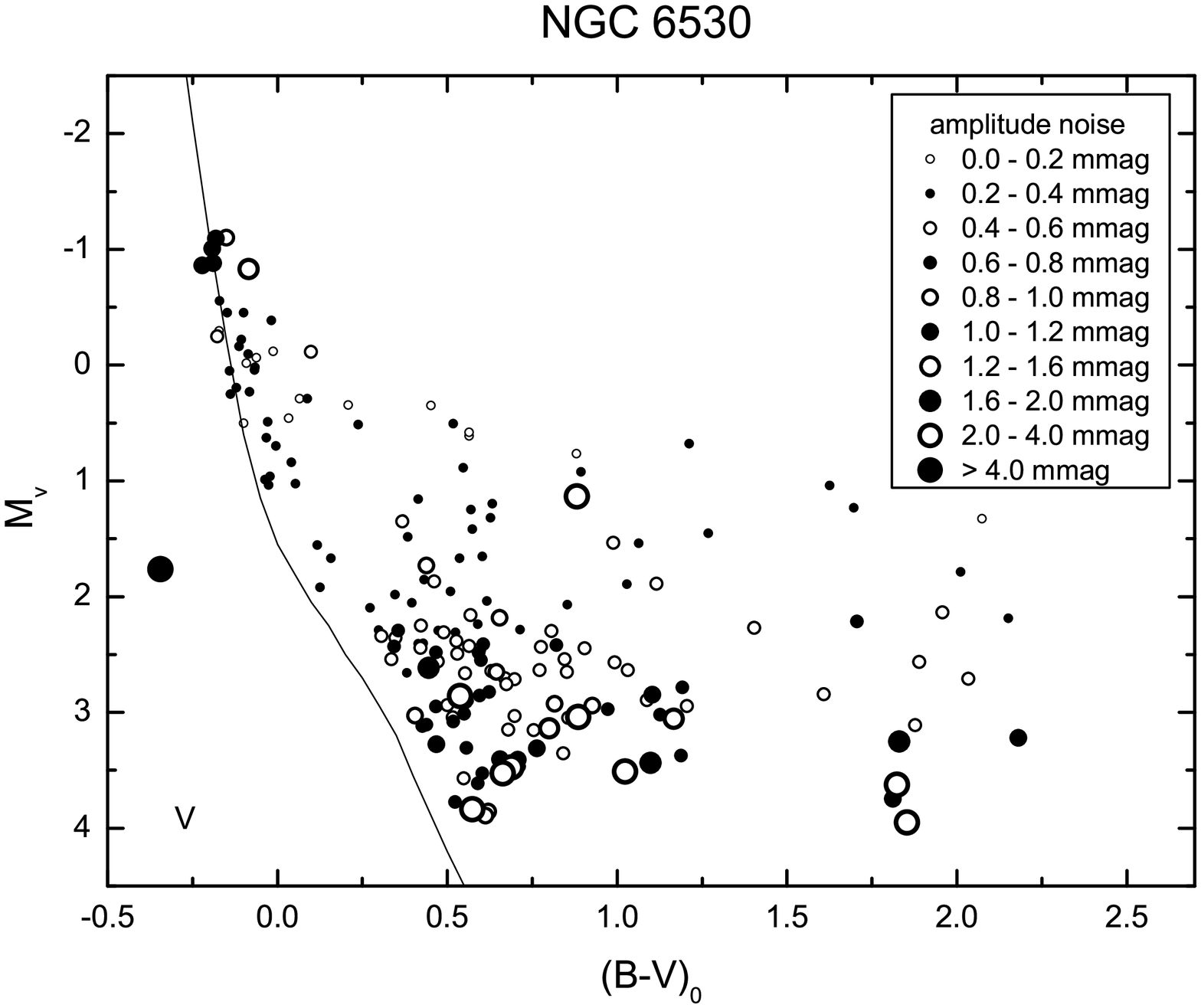}
\includegraphics[width=8.5cm]{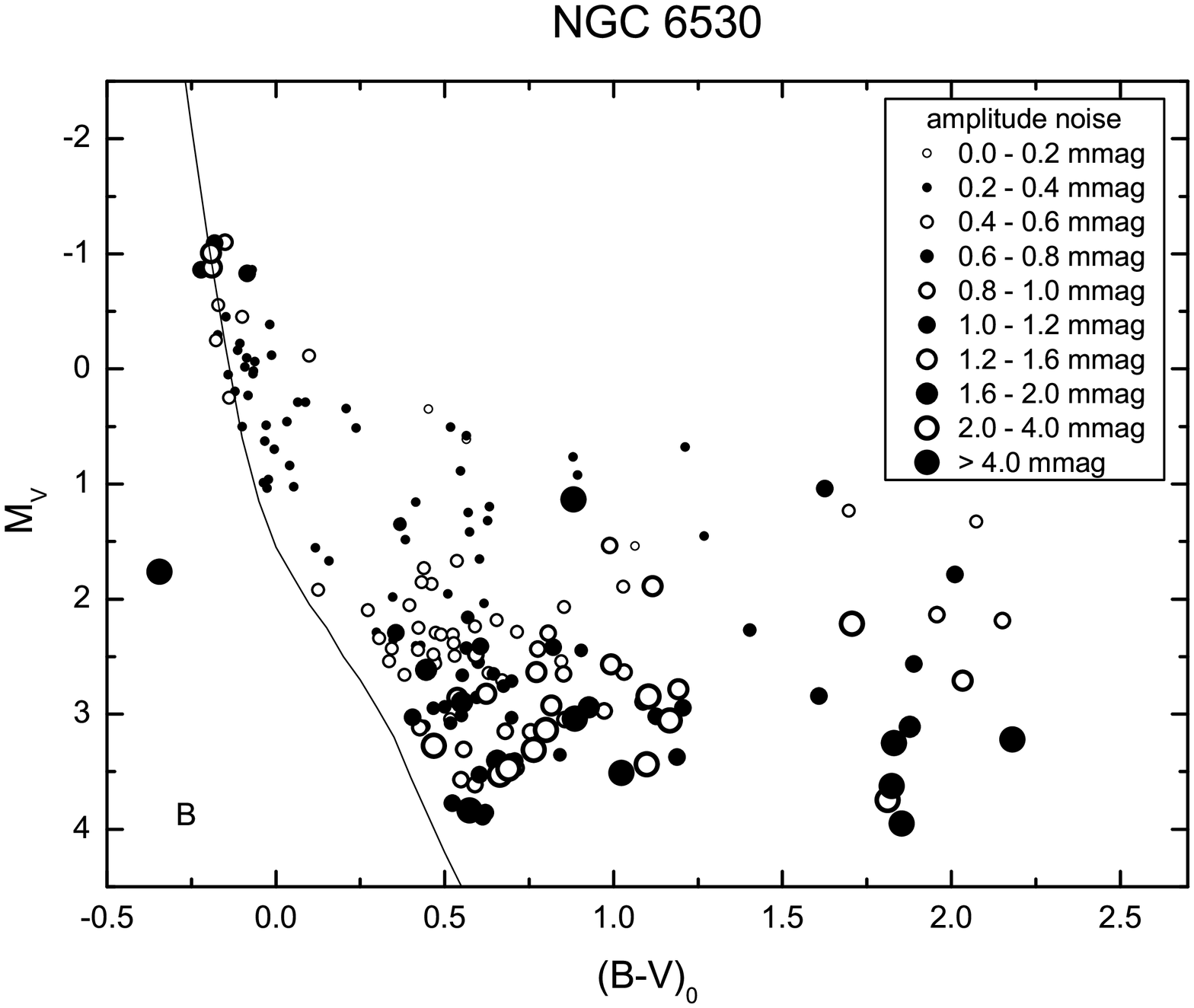}
\caption{HR diagram of all measured stars in the field of NGC 6530, where the stars have different symbols and sizes according to their expected amplitude noise level bins; {\it top:} $V$ filter, {\it bottom:} $B$ filter; the values for the ZAMS (solid line) are taken from Schmidt-Kaler (\cite{sch65}).}
\label{ngc6530-hrd-amp}
\end{figure}

Analogous to IC 4996, the amplitude noise level based on the point-to-point scatter in the time domain has been calculated. Figure \ref{ngc6530-hrd-amp} shows the HR diagrams (top: $V$ and bottom: $B$ filters), where the stars have different symbols and sizes according to their amplitude noise level bins. Please note that again some blue stars are too bright and were partly saturated on the CCD chip. Hence, their number of data points is somewhat lower and their accuracy is worse than for their slightly fainter counterparts.

\section{Conclusions}
CCD photometric time-series of 113 stars in IC 4996 and 194 stars in NGC 6530 have been analyzed with the aim of searching for pulsating \pms members of the clusters using classical Fourier techniques as well as the new concept {\sc SigSpec} (Reegen \cite{ree06}).

40 stars are situated in the instability region of the HR diagram in IC 4996, and hence have been primary candidates in which to search for pulsations. Only two cluster members, IC 4996 37 and IC 4996 40, have been identified as bona fide PMS pulsators, whereas pulsation can only be suspected for IC 4996 46. According to the mid-A spectral types of the two bona fide pulsating PMS stars and their location in the HR diagram, the probability of their membership in the cluster is very high.
Those three stars correspond to only $\la$8\% pulsators within the region of the classical instability strip in IC 4996 down to an amplitude noise level of 0.2 mmag in {\it V} and 0.6 mmag in {\it B} in the Fourier domain. The percentage of detected pulsating PMS stars in the instability region is somewhat lower than is observed for their post-ZAMS counterparts, even if the not optimal data quality is taken into account. However, this finding is still consistent with earlier investigations of pulsating PMS stars in young clusters (e.g., Zwintz et al. \cite{zwi05}).

In NGC 6530 the total number of potential candidates for PMS pulsation of 30 is less than for IC 4996. Still, significantly more pulsating \pms stars have been discovered in NGC 6530. Six stars (NGC 6530 5, NGC 6530 82, NGC 6530 85, NGC 6530 263, NGC 6530 278, and NGC 6530 281) definitely show $\delta$ Scuti-like pulsation periods, while for one star (NGC 6530 288) pulsation can only be suspected. This corresponds to $\sim$24\% pulsating stars in the region of the classical instability strip down to an amplitude noise level of 0.2 mmag in both filters in the Fourier domain. Three of the PMS pulsators in NGC 6530 have been included in the proper motion study by van Altena \& Jones (\cite{alt72}), who report a high probability of cluster membership.
For the other stars, the membership was estimated from their position in the HR diagram and from their location on the sky with respect to the center of the cluster.

Often not much additional information is available for a star, e.g., for most of the objects spectral classification would urgently be needed. For the couple of pulsating stars with known spectral types, mostly emission lines have been reported in the literature, and very often an excess in the infrared was also observed. From stellar evolution theory it is suspected that such young stars are fast rotators, but for nearly all known PMS pulsators no \vsini determination has been performed yet, making it difficult to confirm this concept. Additionally, the detected periodicities due to pulsation always lie on top of longer, sometimes irregular light variations caused by circumstellar material. All these features are characteristics of the Herbig Ae/Be stars indicating their \pms nature.

From the pulsational analysis of the 8 pulsating \pms stars discovered in the two clusters, it is evident that the higher overtone modes seem to have higher amplitudes in \pms stars and hence are easier to detect. This is confirmed by the asteroseismic investigations of other detected PMS pulsators (e.g. Ripepi et al. \cite{rip06}). At the moment, the reason for this behavior can only be speculated about: the amplitudes of fundamental and lower overtone modes seem to be damped by convection, which favor higher overtones.

A total of 25 PMS $\delta$ Scuti-like stars and 10 PMS $\delta$ Scuti-like candidates are known in the literature, including the 10 bona fide and 2 candidate PMS pulsators found in this study. Such a number will allow to probe the instability strip for \pms stars empirically (Zwintz \cite{zwi06}) for the first time, but also supports the ongoing development of non-radial pulsation models for \pms stars.

\begin{acknowledgements}

This project was supported by the Austrian {\it Fonds zur F\"orderung der wissenschaftlichen Forschung} (P14984). Frequency analysis was performed using {\sc Period98} written by M. Sperl (1998) and {\sc SigSpec} developed by P. Reegen (2006).
Use was made of the WEBDA database, developed by J.-C.Mermilliod (Laboratory of Astrophysics of the EPFL, Switzerland) and operated by E. Paunzen at the University of Vienna, Austria.
Special thanks goes to Rafa Garrido and the team of the Sierra Nevada Observatory for carrying out the observations of IC 4996.
Finally, it is a pleasure to acknowledge T. Kallinger, A. Pamyatnykh, and P. Reegen for fruitful discussions on data reduction and analysis.

\end{acknowledgements}


\begin{thebibliography}{}
\bibitem[1985]{alf85}Alfaro, E.J., Delgado, A.J., Garc\'ia-Pelayo, J.M., et al. 1985, \aaps, 59, 441
\bibitem[1972]{bre72}Breger, M. 1972, \apj, 171, 539
\bibitem[1979]{bre79}Breger, M. 1979, \pasp, 91, 5
\bibitem[1993]{bre93}Breger, M., Stich, J., Garrido, R., et al. 1993, \aap, 271, 482
\bibitem[1998]{bre98}Breger, M., \& Pamyatnykh, A. 1998, \aap, 332, 958
\bibitem[2000]{bre00}Breger, M. 2000, ASP Conf. Ser., 210, 3
\bibitem[1981]{chi81}Chini, R., \& Neckel, Th. 1981, \aap, 102, 171
\bibitem[1975]{dee75}Deeming, T.J. 1975, \apss, 36, 137
\bibitem[1998]{del98}Delgado, A.J., Alfaro, E.J., \& Moitinho, A. 1998, \aj, 116, 1801
\bibitem[1999]{del99}Delgado, A.J., Miranda, L.F., \& Alfaro, E. 1999, \aj, 118, 1759
\bibitem[1997]{don97}Donati, J.-F., Semel, M., Carter, B.D., et al. 1997, \aap, 291, 658
\bibitem[1984]{fin84}Finkenzeller, U., \& Mundt, R. 1984, \aaps, 55, 109
\bibitem[1960]{her60}Herbig, G.H. 1960, \apjs, 4, 337
\bibitem[2005]{kal05}Kallinger, T. 2005, Comm. Ast., 146, 45
\bibitem[1977]{kil77}Kilambi, G.C. 1977, \mnras, 178, 423
\bibitem[1995]{kur95}Kurtz, D., \& Marang, F. 1995, \mnras, 276, 191
\bibitem[1997]{kus97}Kuschnig, R., Weiss, W.W., Gruber, et al. 1997, \aap, 328, 544
\bibitem[1990]{loz90}Lozinskaya, T.A., \& Repin, S.V. 1990, \sovast, 34, 580
\bibitem[1998]{mar98}Marconi, M., \& Palla, F. 1998, \aj, 507, L141
\bibitem[2000]{mar00}Marconi, M., Ripepi, V., Alcal\'a, J.M., et al. 2000, \aap, 355, L35
\bibitem[1990]{mcc90}McCall, M.L., Richer, M.G., \& Visvanathan, N. 1990, \apj, 357, 502
\bibitem[1999]{mon99}Montgomery, M. 1999, Delta Scuti Newsletter, 13, 28
\bibitem[2000]{pam00}Pamyatnykh, A.A. 2000, ASP Conf. Ser., 210, 215
\bibitem[1998]{ree98}Reed, B.C. 1998, \jrasc, 92, 36
\bibitem[2006]{ree06}Reegen, P. 2006, \mnras, submitted
\bibitem[2006]{rip06}Ripepi, V., Bernabei, S., Marconi, M., et al. 2006, \aap, accepted
\bibitem[1965]{sch65}Schmidt-Kaler, Th. 1965, in Landolt-B\"ornstein: Numerical data and functional relationships in science and technology, ed. H.H. Voigt (Berlin: Springer Verlag), group VI, vol. I, 284
\bibitem[1998]{spe98}Sperl, M. 1998, Co. Ast., 111, 1
\bibitem[2000]{sun00}Sung, H., Chun, M., \& Bessel, M. 2000, \aj, 120, 333
\bibitem[1930]{tru30}Trumpler, R.J. 1930, Lick Obs. Bull, 14, 154
\bibitem[1972]{alt72}van Altena, W.F., \& Jones, B.F. 1972, \aap, 20, 425
\bibitem[1997]{van97}van den Ancker, M.E., Th\'e, P.S., Feinstein, A., et al. 1997, \aaps 123, 63
\bibitem[1996]{van96}Vansevi\u{c}ius, V., Brid\u{z}ius, A., Pu\u{c}inskas, A., et al. 1996, Baltic Astronomy, vol. 5, 539
\bibitem[1991]{voi91}Voigt, H.H. 1991, in Abriss der Astronomie. (Mannheim/Wien/Z\"urich: B.I. Wissenschaftsverlag)
\bibitem[2005]{zwi05}Zwintz, K., Marconi, M., Reegen, P., et al. 2005, \mnras, 357, 345
\bibitem[2006]{zwi06}Zwintz, K. 2006, \aj, in preparation
\end{thebibliography}
\end{document}